\documentclass[pdflatex,sn-mathphys-num]{sn-jnl}

\usepackage{graphicx}%
\usepackage{multirow}%
\usepackage{amsmath,amssymb,amsfonts}%
\usepackage{amsthm}%
\usepackage{mathrsfs}%
\usepackage[title]{appendix}%
\usepackage{xcolor}%
\usepackage{textcomp}%
\usepackage{manyfoot}%
\usepackage{booktabs}%
\usepackage{algorithm}%
\usepackage{algorithmicx}%
\usepackage{algpseudocode}%
\usepackage{listings}%
\usepackage{bm}
\usepackage{physics}
\usepackage{comment}
\usepackage{siunitx}

\theoremstyle{thmstyleone}%
\theoremstyle{thmstyletwo}%

\theoremstyle{thmstylethree}%

\begin{document}

\title{High-Rate Four Photon Subtraction from Squeezed Vacuum: Preparing Cat State for Optical Quantum Computation}


\author*[1,2]{\fnm{Mamoru} \sur{Endo}}\email{endo@ap.t.u-tokyo.ac.jp}
\author[1]{\fnm{Takefumi} \sur{Nomura}}
\author[1]{\fnm{Tatsuki} \sur{Sonoyama}}
\author[1]{\fnm{Kazuma} \sur{Takahashi}}
\author[3]{\fnm{Sachiko} \sur{Takasu}}
\author[3,4]{\fnm{Daiji} \sur{Fukuda}}
\author[5]{\fnm{Takahiro} \sur{Kashiwazaki}}
\author[5]{\fnm{Asuka} \sur{Inoue}}
\author[5]{\fnm{Takeshi} \sur{Umeki}}
\author[1,6,7,8]{\fnm{Rajveer} \sur{Nehra}}
\author[9]{\fnm{Petr} \sur{Marek}}
\author[9]{\fnm{Radim} \sur{Filip}}
\author[1,2]{\fnm{Kan} \sur{Takase}}
\author[1,2]{\fnm{Warit} \sur{Asavanant}}
\author*[1,2]{\fnm{Akira} \sur{Furusawa}}\email{akiraf@ap.t.u-tokyo.ac.jp}

\affil*[1]{\orgdiv{Department of Applied Physics, School of Engineering}, \orgname{The University of Tokyo}, \orgaddress{\street{7-3-1 Hongo}, \city{Bunkyo}, \state{Tokyo}, \postcode{113-8656}, \country{Japan}}}
\affil[2]{\orgdiv{Optical Quantum Computing Research Team}, \orgname{RIKEN Center for Quantum Computing}, \orgaddress{\street{2-1 Hirosawa}, \city{Wako}, \state{Saitama}, \postcode{351-0198}, \country{Japan}}}
\affil[3]{\orgdiv{National Institute of Advanced Industrial Science and Technology}, \orgaddress{\street{1-1-1 Umezono}, \city{Tsukuba}, \state{Ibaraki}, \postcode{305-8563}, \country{Japan}}}
\affil[4]{\orgdiv{AIST-UTokyo Advanced Operando-Measurement Technology Open Innovation Laboratory}, \orgaddress{\street{1-1-1 Umezono}, \city{Tsukuba}, \state{Ibaraki}, \postcode{305-8563}, \country{Japan}}}
\affil[5]{\orgdiv{NTT Device Technology Labs}, \orgname{NTT Corporation}, \orgaddress{\street{3-1 Morinosato Wakamiya}, \city{Atsugi}, \state{Kanagawa}, \postcode{243-0198}, \country{Japan}}}
\affil[6]{\orgdiv{Department of Electrical and Computer Engineering}, \orgname{University of Massachusetts Amherst}, \orgaddress{\street{100 Natural Resources Rd}, \city{Amherst}, \state{Massachusetts}, \postcode{01003}, \country{USA}}}
\affil[7]{\orgdiv{Department of Physics}, \orgname{University of Massachusetts Amherst}, \orgaddress{\street{100 Natural Resources Rd}, \city{Amherst}, \state{Massachusetts}, \postcode{01003}, \country{USA}}}
\affil[8]{\orgdiv{College of Information and Computer Science}, \orgname{University of Massachusetts Amherst}, \orgaddress{\street{100 Natural Resources Rd}, \city{Amherst}, \state{Massachusetts}, \postcode{01003}, \country{USA}}}
\affil[9]{\orgdiv{Department of Optics}, \orgname{Palacky University}, \orgaddress{\street{17. listopadu 1192/12}, \city{Olomouc}, \postcode{77146}, \country{Czech Republic}}}


\abstract{Generating logical qubits, essential for error detection and correction in quantum computation, remains a critical challenge in continuous-variable (CV) optical quantum information processing. The Gottesman-Kitaev-Preskill (GKP) code is a leading candidate for logical qubits, and its generation requires large-amplitude coherent state superpositions---Schr\"{o}dinger cat states. However, experimentally producing these resource states has been hindered in the optical domain by technical challenges. The photon subtraction method, a standard approach for generating cat states using a squeezed vacuum and a photon number-resolving detector, has proven difficult to scale to multi-photon operations. While the amplitude of the generated cat states increases with the number of subtracted photons, limitations in the generation rate have restricted the maximum photon subtraction to $n=3$ for over a decade.
In this work, we demonstrate high-rate photon subtraction of up to four photons from a squeezed vacuum with picosecond wavepackets generated by a broadband optical parametric amplifier. Using a Ti-Au superconducting-transition-edge sensor, we achieve high-speed, high-resolution photon number discrimination. The resulting states exhibit Wigner function negativity without loss correction, and their quantum coherence is verified through off-diagonal density matrix elements in CV representation.
These results overcome long-standing limitations in multi-photon operations, providing a critical foundation for generating quantum resources essential for fault-tolerant quantum computing and advancing ultrafast optical quantum processors.}

\keywords{Optical logical qubit, Shr\"{o}dinger cat state, Photon subtraction, Transition-edge sensor, Optical parametric amplifier}

\maketitle

\section{Introduction}\label{sec1}
Light, with its exceptionally high frequency in the hundreds of terahertz, is not only a fundamental carrier of information for ultra-fast communication and data processing but also a promising platform for quantum technologies. Its ability to maintain quantum coherence under ambient conditions has positioned it as a key candidate for quantum computing \cite{Lloyd1999,Takeda2019,Asavanant2019,Larsen2019, Aghaee2025}, quantum sensing \cite{Sudbeck2020,Anisimov2010}, and quantum communication \cite{Couteau2023}. The realization of ultra-fast quantum processors by encoding quantum information into the continuous-variable (CV) quadrature-phase amplitudes of the electromagnetic field is a particularly promising direction.

A major challenge in realizing fault-tolerant quantum computation, however, lies in the generation of logical qubits, which are necessary for error correction. In optical systems, the Gottesman-Kitaev-Preskill (GKP) qubit, characterized by a comb-like wavefunction structure, has emerged as a strong candidate for logical qubits \cite{Gottesman2001}. GKP qubits offer a significant advantage in that their fundamental operations, such as Clifford gates, can be implemented using only linear optics. Furthermore, universal quantum computation becomes achievable with these operations combined with ancilla GKP qubits \cite{Baragiola2019,Konno2021,Aghaee2025}. Therefore, the successful realization of GKP qubits in optical systems is a critical step toward fault-tolerant universal quantum computing.

Several theoretical proposals have outlined methods to generate GKP qubits \cite{Vasconcelos2010,Weigand2018,Eaton2019, Tzitrin2020,Takase2023}, notably by using Schr\"{o}dinger cat states---quantum superpositions of coherent states \cite{Cochrane1999,Ralph2003}. These proposals involve the preparation of multiple cat states, which are then interfered with and processed using feed-forward techniques. While proof-of-principle experiments have demonstrated this approach with two cat states \cite{Konno2024}, significant challenges remain in scaling the method. Producing high-quality GKP qubits requires cat states with large amplitudes, but their generation has proven difficult. Furthermore, the need for multiple cat states increases the complexity of interference and feed-forward operations, making the overall generation rate of GKP qubits prohibitively low. 

A primary limitation arises from the probabilistic nature of the photon subtraction technique commonly used to generate cat states in optical systems \cite{Dakna1997}. This method involves tapping a portion of a squeezed vacuum field via a beam splitter and detecting photons in one output arm. The detection of $n$ photons creates a quantum state in the other output that approximates a Schr\"{o}dinger cat state. The amplitude of this state is proportional to the square root of the number of subtracted photons. While single-photon subtraction has been extensively studied \cite{Ourjoumtsev2006,Neergaard-Nielsen2008,Asavanant2017,Takase2021}, generating large amplitude cat states requires multiple-photon subtraction. Experimental constraints such as photon number-resolving detection capabilities, optical losses, and phase stability have limited photon subtraction to three photons for over a decade \cite{Gerrits2010,Endo2023}. Additionally, achieving high-quality states often necessitates spectral narrowing of the light source, which reduces the pulse repetition rate and is counterproductive to the high-speed requirements of quantum computation \cite{Yukawa2013}. Previous studies have observed Wigner function negativity---a signature of non-classicality and is necessary for quantum computation \cite{Mari2012}---at rates no larger than ten counts per second (cps) and going as low as thousandths of cps already for three-photon subtraction \cite{Gerrits2010, Yukawa2013, Endo2023}, making higher photon subtraction rates impractical. 

In this study, we overcome these limitations by employing a broadband waveguide-based optical parametric amplifier \cite{Kashiwazaki2023} pumped with picosecond pulses to generate pulsed squeezed vacuum fields with a wavepacket duration of about \SI{10}{ps}. Using a high-speed, high-energy-resolution photon number-resolving detector (PNRD) based on a Ti-Au superconducting transition-edge sensor (TES) \cite{Hattori2022}, we successfully demonstrated up to four-photon subtraction. The generation rate for three-photon subtraction was approximately \SI{200}{cps}, and even for four-photon subtraction, the rate was \SI{1.5}{cps}. To the best of the authors' knowledge, there are no previous reports of nonclassical state generation experiments involving more than three photons using the heralded method. 
The resulting cat states exhibit evidence of non-classicality, including Wigner function negativity without loss correction and parity changes corresponding to the number of subtracted photons.  Furthermore, the off-diagonal elements of the density matrix in the quadrature basis complementarily confirm the generation of coherent superpositions rather than mixed coherent states. The experimentally generated states closely match theoretical simulations in qualitative and quantitative aspects.

The method established in this study is highly versatile and can be readily extended to next-generation quantum state synthesis techniques, such as generalized photon subtraction \cite{Takase2021}. This approach is expected to directly lead to the realization of ultra-fast fault-tolerant optical quantum computers, fully leveraging the high-speed nature of light.
\begin{figure*}
\centering
\includegraphics[width=\textwidth]{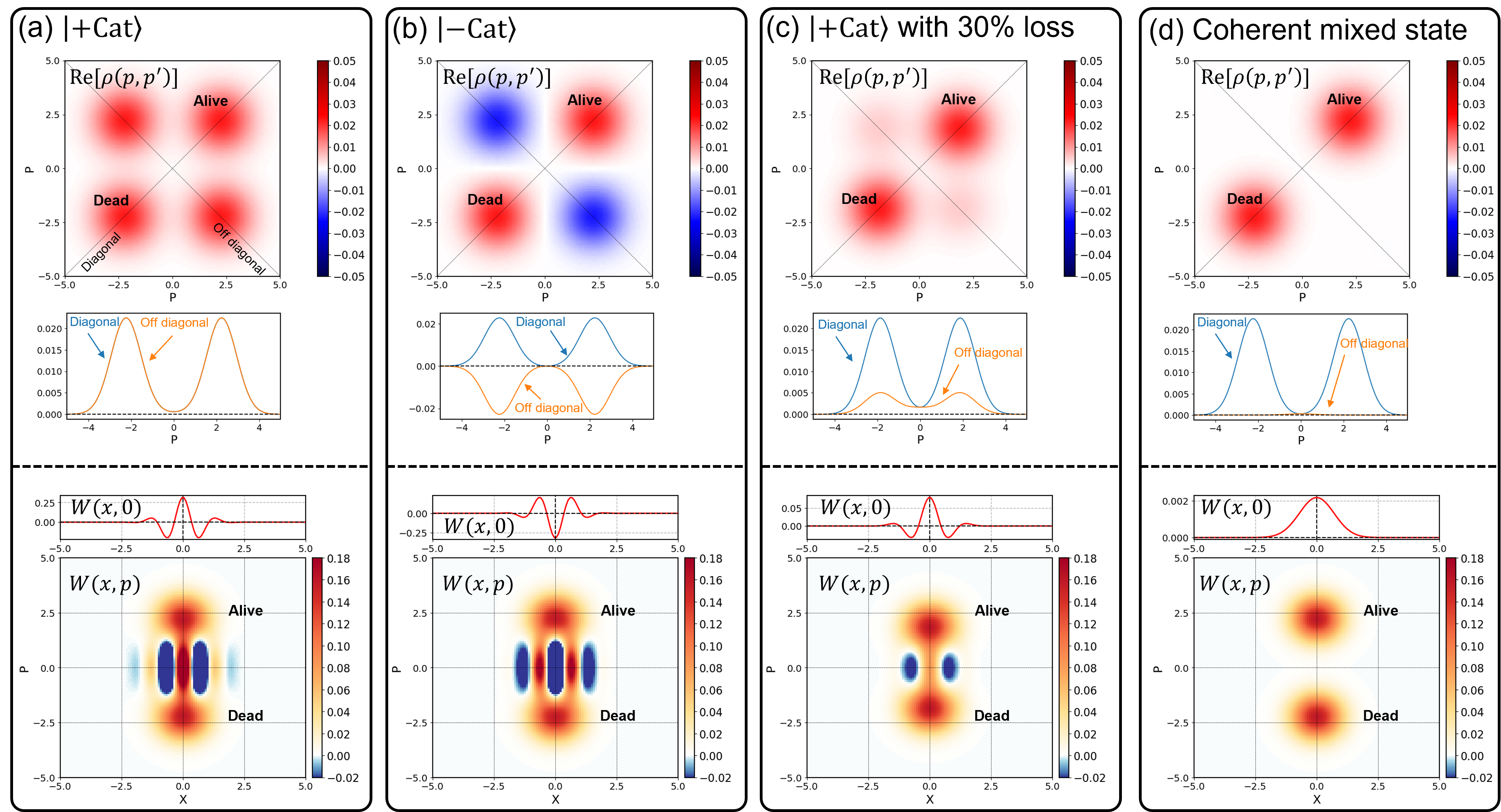}
\caption{Simulation examples of the density matrices in the momentum bases $\mathrm{Re}[\rho(p,p')]$ and Wigner functions $W(x,p)$ corresponding to the quantum state. The plots shown below the density matrix heatmaps represent the diagonal (blue, $\mathrm{Re}[\rho(p,p)]$) and off-diagonal (orange, $\mathrm{Re}[\rho(p,-p)]$) components. The plots shown above the Wigner function heatmaps represent the cross-section $W(x,0)$ at $p=0$. (a) Even cat state $\ket{\mathrm{+Cat}}\propto\ket{\alpha}+\ket{-\alpha}$. (b) Odd cat state $\ket{\mathrm{-Cat}}\propto\ket{\alpha}-\ket{-\alpha}$ (c) Even cat state $\ket{\mathrm{+Cat}}$ with 30\% loss. (d) Coherent mixed state $\dyad{\alpha}{\alpha}+\dyad{-\alpha}{-\alpha}$. All plots were simulated with $\alpha=2.5i$. Note that in
(a), the diagonal and off-diagonal lines of the density matrix are fully overlapped.}\label{fig:catstates}.
\end{figure*}

\begin{figure}
\centering
\includegraphics[width=0.6\textwidth]{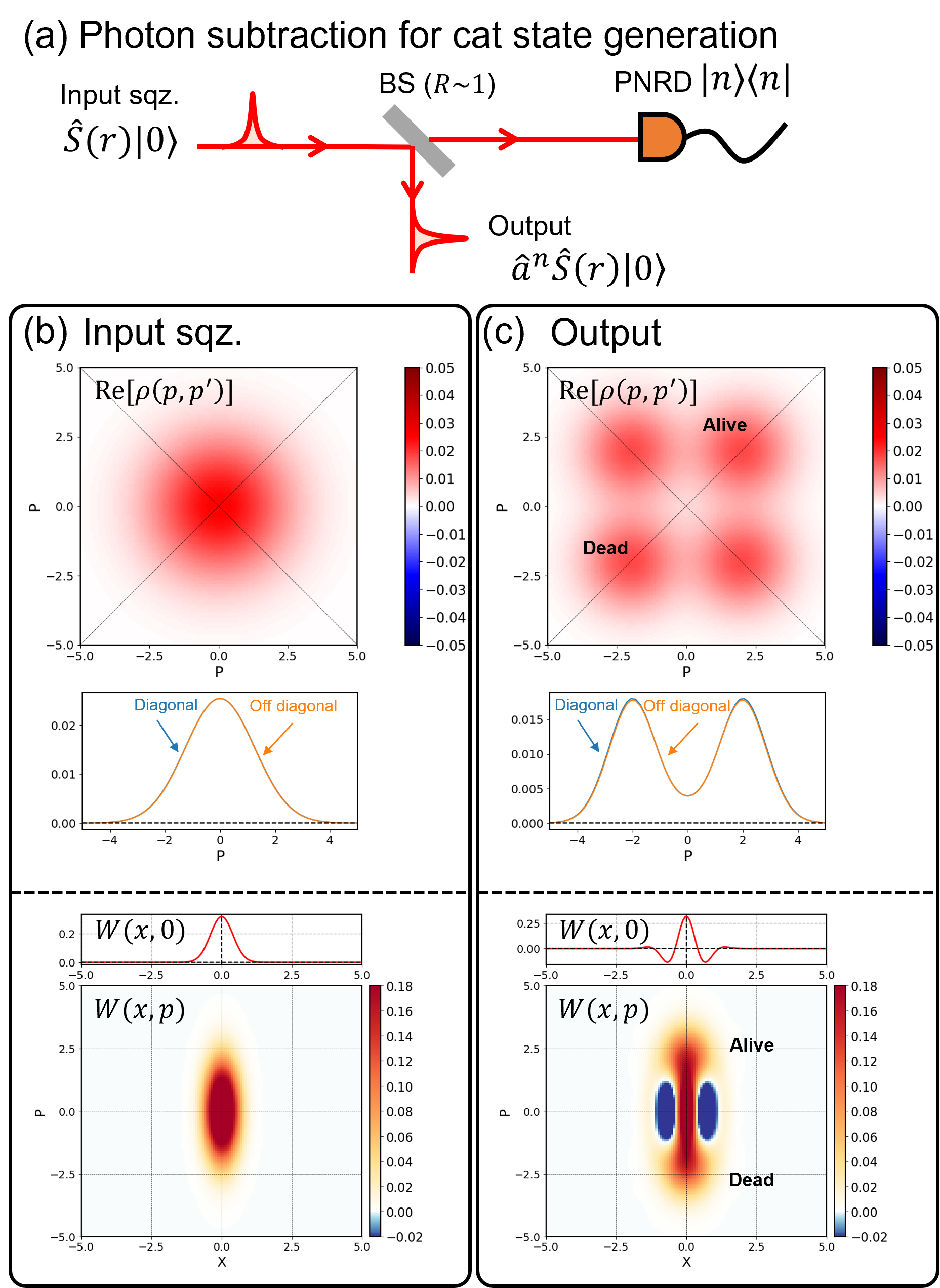}
\caption{(a) Conceptual diagrams of cat state generation by photon subtraction from squeezed vacuum. Simulated density matrices in the momentum bases (top, $\rho(p,p')$) and Wigner functions (bottom, $W(x,p)$) of (b) input squeezed vacuum (input sqz. $\hat{S}(0.576)\ket{0}$, that is, squeezing level of \SI{5}{dB}), and (c) four-photon-subtracted squeezed vacuum $\propto\hat{a}^4\hat{S}(0.576)\ket{0}$ with $R=0.78$. Note that, in (b) and (c), the diagonal and off-diagonal lines are overlapped. }\label{fig:photonsubtraction}
\end{figure}

\section{Results}\label{sec2}
\subsection{Cat states}
Coherent state superpositions hold particular significance in quantum optics. These states are often referred to as Schr\"{o}dinger cat states, in reference to the famous Gedankenexperiment \cite{Schrodinger1935}. In this analogy, the states of an “alive cat” and a “dead cat” correspond to the coherent states with amplitudes of $\alpha$ and $-\alpha$, respectively. The state vectors of the cat state are expressed as $\ket{\pm\mathrm{Cat}}\propto\ket{\alpha}\pm\ket{-\alpha}$, where $\ket{+\mathrm{Cat}}$ means even cat and $\ket{-\mathrm{Cat}}$ means odd cat. The corresponding density matrices are $\rho_{\pm\mathrm{Cat}}\propto\dyad{\alpha}{\alpha}+\dyad{-\alpha}{-\alpha}\pm\dyad{\alpha}{-\alpha}\pm\dyad{-\alpha}{\alpha}$. The components ($\dyad{\alpha}{-\alpha},\dyad{-\alpha}{\alpha}$) represent quantum interference of the coherent states. In contrast, the density matrix of a classical mixture of coherent states is given by $\rho_\mathrm{mixed}\propto\dyad{\alpha}{\alpha}+\dyad{-\alpha}{-\alpha}$, which lacks the interference of the coherent states. Cat states are valuable not only as standalone quantum error-correcting codes but also as resources for generating more robust codes such as GKP qubits \cite{Vasconcelos2010,Weigand2018,Tzitrin2020, Takase2023}.

Several CV representations exist for describing quantum states of light, however, a combination of the density matrix $\rho(p,p')=\mel{p}{\rho}{p'}$ and Wigner function $W(x,p)$ in the CV formalism are particularly illustrative for the cat-state interference.  Both of these pictures contain all properties of the state, but they make them visible in different ways. While the density matrix directly shows the correlations between different momentum states, the Wigner function, defined as
\begin{align}
W(x,p)=\frac{1}{\pi\hbar}\int_{-\infty}^{\infty}e^{2ip'x/\hbar}\mel{p+p'}{\rho}{p-p'}dp',
\end{align}
applies a partial Fourier transformation to represent the relation between position and momentum better instead. Figure~\ref{fig:catstates} presents examples of two types of cat states ((a) even or (b) odd cats), (c) a degraded even cat state with 30\% loss, and (d) a coherent mixed state, where all plots are simulated with $\alpha = 2.5i$, corresponding to a superposition of the opposite momenta in the coherent states. The density matrices and Wigner functions corresponding to these states complementarily highlight their key differences, particularly in the off-diagonal elements, which signify quantum interference.

For instance, all the depicted density matrices exhibit two peaks in their diagonal components (indicated as “Alive” and “Dead” in the heatmaps), corresponding to the amplitudes of $\alpha$ and $-\alpha$, respectively. The primary distinction lies in the off-diagonal components $\rho(p, -p)=\mel{p}{\rho}{-p}$ reflecting quantum interference of momentum eigenstates inside the cat. Cat states exhibit pronounced off-diagonal elements, whose sign reflects the parity of the cat state (even or odd). In contrast, these off-diagonal elements diminish under loss, as evident in the degraded cat state, and are absent in the case of coherent mixed states.

The Wigner function $W(x,p)$ is another common tool for visualizing quantum states. As shown in the figures, cat states exhibit interference fringes in $W(x, 0)$ near the origin of the phase space, which are a hallmark of quantum coherence. The sign of the cat state’s parity is also evident in the value of the Wigner function at the origin. Notably, the presence of negative values in the Wigner function is a strong indicator of the state’s non-classical nature. Achieving quantum states with such negativity in the Wigner function without applying loss correction remains a key experimental goal, as it signifies a high degree of non-classicality. We present both representations interconnected by a Fourier transform simultaneously to understand the nonclassical aspects of momentum interference.

\subsection{Photon subtraction}
Approximate cat states can be generated by photon subtraction \cite{Dakna1997} as shown in Fig.~\ref{fig:photonsubtraction}(a). A squeezed vacuum state $\hat{S}(r)\ket{0}$ is deterministically generated via an optical parametric process and split into two modes by a high-reflectivity beam splitter (BS) with reflectivity of $R$ (usually $R\sim1$), where $\hat{S}(r)$ and $r$ are a squeezing operator and a squeezing parameter, respectively. The transmitted mode is referred to as the idler mode, and the reflected mode as the signal mode.

Next, photon-number-resolving detection is performed on the idler mode. When $n$ photons are detected, the signal mode collapses into the state $\hat{a}^n\hat{S}(r)\ket{0}$ ($\hat{a}$ is an annihilation operator), which serves as an approximation of a cat state. Notably, the parity (even or odd) of the generated cat state matches that of $n$, as the squeezed vacuum state consists of even-photon-number superpositions. This approach is commonly referred to as the photon subtraction method, as it resembles subtracting photons from a squeezed vacuum state. 
Figure~\ref{fig:photonsubtraction} (b) shows the relevant density matrix $\mathrm{Re}[\rho(p,p')]$ and Wigner function $W(x,p)$ of a squeezed vacuum state with $r=0.567$ (labeled ``input sqz.''). The density matrix contains no off-diagonal elements, and the Wigner function exhibits no negative values. In Fig.~\ref{fig:photonsubtraction} (c), the density matrix and Wigner function of the simulated state generated by tapping the initial squeezed vacuum with a beam splitter ($R=0.81$) and detecting four photons in the idler mode are presented. The density matrix reveals pronounced off-diagonal elements. The Wigner function shows negative values, consistent with non-classicality, which is not achievable by the input squeezed state. Note that, for $n=4$, the off-diagonal components of the density matrix are positive, and the value of the Wigner function at the origin is also positive.

\subsection{Experimental apparatus}
\begin{figure}
\centering
\includegraphics[width=\textwidth]{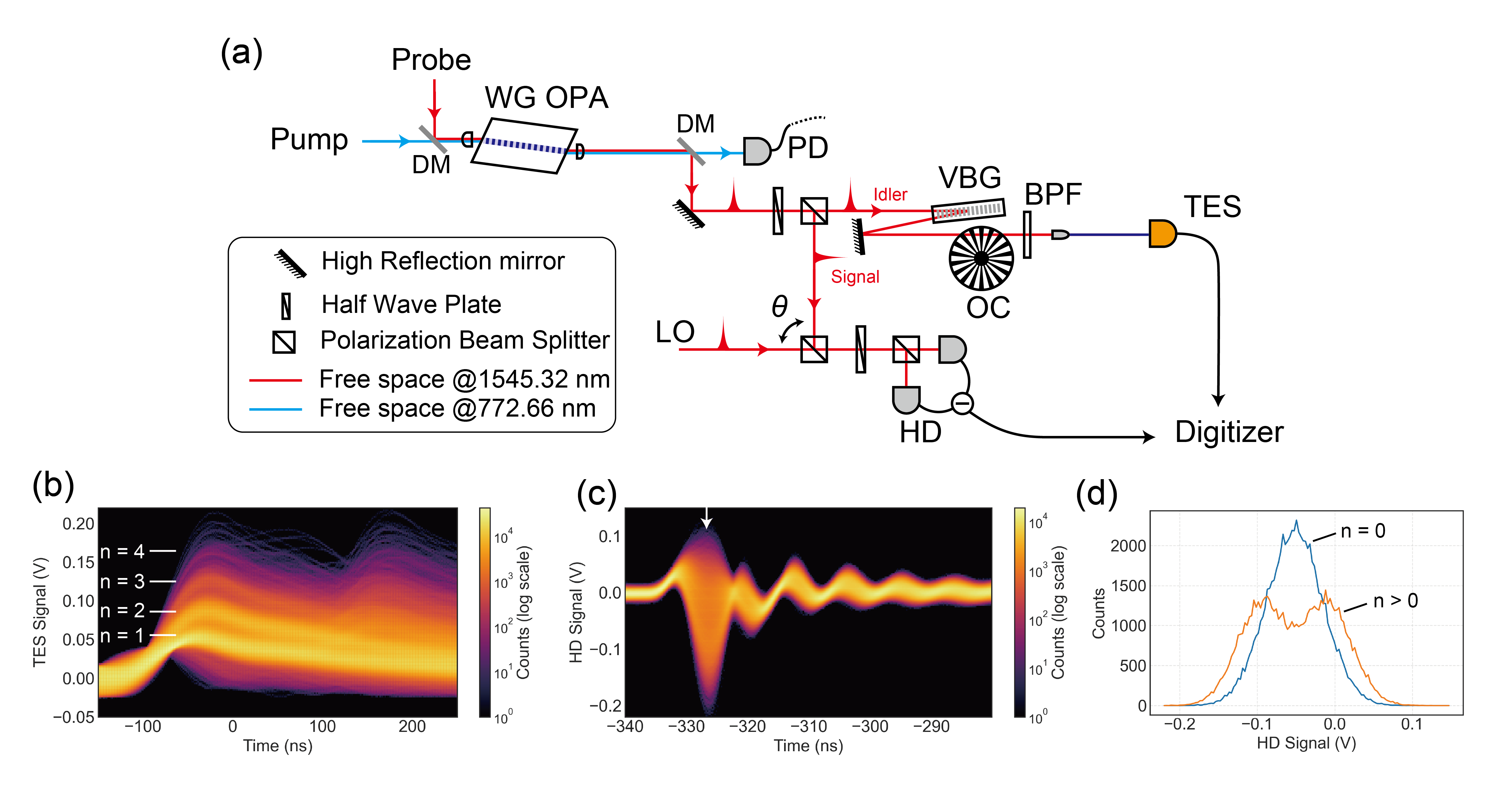}
\caption{(a) Experimental apparatus for the 4-photon subtraction. DM, dichroic mirror; WG OPA, waveguide optical parametric amplifier; PD, photodetector; VBG, volume-Bragg grating; BPF, optical bandpass filter; OC, optical chopper; TES, Ti-Au transition-edge sensor; LO, local oscillator; HD, homodyne detector. (b) shows a heatmap of TES signals over approximately 70,000 frames, where signals corresponding to $n=1,2,3,4$ photons are observed. (c) shows a heatmap of HD signals at a fixed LO phase over approximately 70,000 frames when photon detection was performed using the TES ($n > 0$).  The quadrature-phase amplitudes were calculated based on the voltage values at the timestamps indicated by the downward arrow in the figure. (d) shows voltage histograms of the HD signal at the downward arrow (blue: $n=0$, orange: $n > 0$).}\label{setup}
\end{figure}

The experimental apparatus is illustrated in Fig.~\ref{setup} (a). For more details (detailed experimental setup, phase locking scheme, loss budget), please refer to the supplement and our previous paper \cite{Endo2023, supplement}. A 10 mm-long waveguide-type periodically-poled lithium niobate crystal was pumped by a laser with a central wavelength of \SI{772.66}{nm}, a pulse width of \SI{10}{ps}, and a repetition rate of \SI{5}{MHz}, generating a pulsed squeezed vacuum field at a central wavelength of \SI{1545.32}{nm}. 

Since this experiment is phase-sensitive, a probe beam with a wavelength of \SI{1545.32}{nm} was introduced coaxially through a dichroic mirror (DM) to measure the parametric gain in the waveguide via a photodiode (PD), thereby referencing the phase of the squeezed vacuum field to the probe beam. The squeezed vacuum is split into two optical paths by a BS composed of a half wave plate and a polarization beam splitter (PBS). When measuring the input quantum state, the reflectivity was set to $R=1$, and for the photon subtraction experiment, the reflectivity was set to $R=0.81$. The transmitted light is referred to as the Idler, while the reflected light is referred to as the Signal. The Idler light passes through a wavelength filters (volume Bragg grating: VBG), is coupled into a single-mode fiber, and is detected by a Ti-Au TES-type PNRD installed in an adiabatic demagnetization refrigerator and the TES is cooled down to \SI{280}{mK} (see Methods). The signal is read out using a superconducting quantum interference device and acquired by a high-speed digitizer. The efficiency on the Idler, including the TES efficiency, coupling efficiency, and other optical losses, is approximately 40\%.

On the Signal side, homodyne detection is performed by combining the signal light with a local oscillator (LO) beam and using a home-made balanced photodetector consisting of a pair of high-efficiency photodiodes and a low-noise transimpedance amplifier. The temporal waveform of the LO beam is optimized using a waveform shaper. The phase of the homodyne measurement can be determined by the interference signal between the probe beam and the LO beam, and it was locked to any desired angle using a waveguide phase modulator (not shown). The results of the homodyne measurement are acquired by the same digitizer.

The digitizer measures the signal waveforms from both the TES and the HD, and the signals are analyzed by the FPGA embedded in the digitizer. Figure~\ref{setup} (b) shows a heatmap of TES signals over approximately 70,000 frames, where signals corresponding to $n=1,2,3,4$ photons are observed. Figure~\ref{setup} (c) shows a heatmap of HD signals over approximately 70,000 frames when photon detection was performed. The quadrature-phase amplitudes were calculated based on the voltage values at the timestamps indicated by the downward arrows. Examples of voltage histograms are shown in Fig.~\ref{setup} (d), where blue and orange lines represent the HD signals, when $n=0$ and $n>0$, respectively. 
The HD signal when the signal light is blocked is also recorded as the shot noise level, and this information is used to normalize the voltage values of the homodyne measurement. Note that the data shown in Fig.~\ref{setup} (b) to (d) are not used for actual tomography. All the homodyne data used for the tomography can be found in the supplementary information \cite{supplement}.

In this experiment, the probe beam used for phase referencing is chopped at a frequency of 500 Hz, and the phase locking of each part of the system is achieved using a sample-and-hold method. The phase is locked when the probe beam is ON, and the state is held when the probe beam is OFF. Photon detection with the TES and homodyne measurements are performed when the probe is OFF to prevent the classical probe light from interfering with the homodyne measurement. Additionally, an optical chopper (OC) is placed before the optical fiber to prevent the probe light from saturating the TES output.

\section{Discussions}\label{sec4}
\subsection{Homodyne measurement and quantum state tomography}
We performed homodyne measurements at various phases ($\theta =-45^\circ$, $-22.5^\circ$, $0^\circ$, $22.5^\circ$, $45^\circ$, $90^\circ$, where $\theta=0^\circ$ and $\theta=90^\circ$
correspond $x$ and $p$, respectively). The measured quadrature values and their histograms can be found in the appendix. We used the resulting dataset to reconstruct the density matrix on the photon-number basis ($\rho_{n,m}=\mel{n}{\rho}{m}$) via the maximum likelihood method \cite{Lvovsky2009}. Then we calculate photon-number distributions, density matrices in the momentum basis, and Wigner functions. Note that density matrices on a quadrature basis can efficiently be reconstructed from the homodyne results by the machine learning method \cite{Fedotova2023}. It is important to note that in this study, no loss correction was applied during the reconstruction of the density matrix or other parameters, including optical losses, mode mismatch in homodyne detection, or detector efficiency.

\subsection{Simulations}
The parameters required for the simulations were determined based on actual experimental conditions (i.e., they were not obtained by fitting to experimental data). The loss of the waveguide OPA was 0.05, the reflectivity of the beam splitter was 0.81, the loss on the idler side was 0.6 (as mentioned earlier), and the loss on the signal side, including losses from optical elements and the homodyne detector (e.g., spatial and temporal mode mismatches and the quantum efficiency of the photodiode), was 0.15. Additionally, the input squeezing level, assuming no losses, was 6.5 dB. Based on these parameters, subsequent simulations were performed using the Strawberryfields quantum optics library \cite{Killoran2019,Bromley2020}. Throughout this paper, $\hbar = 1$ is assumed.

\subsection{Photon number distributions and count rates}
\begin{figure}
\centering
\includegraphics[width=0.6\textwidth]{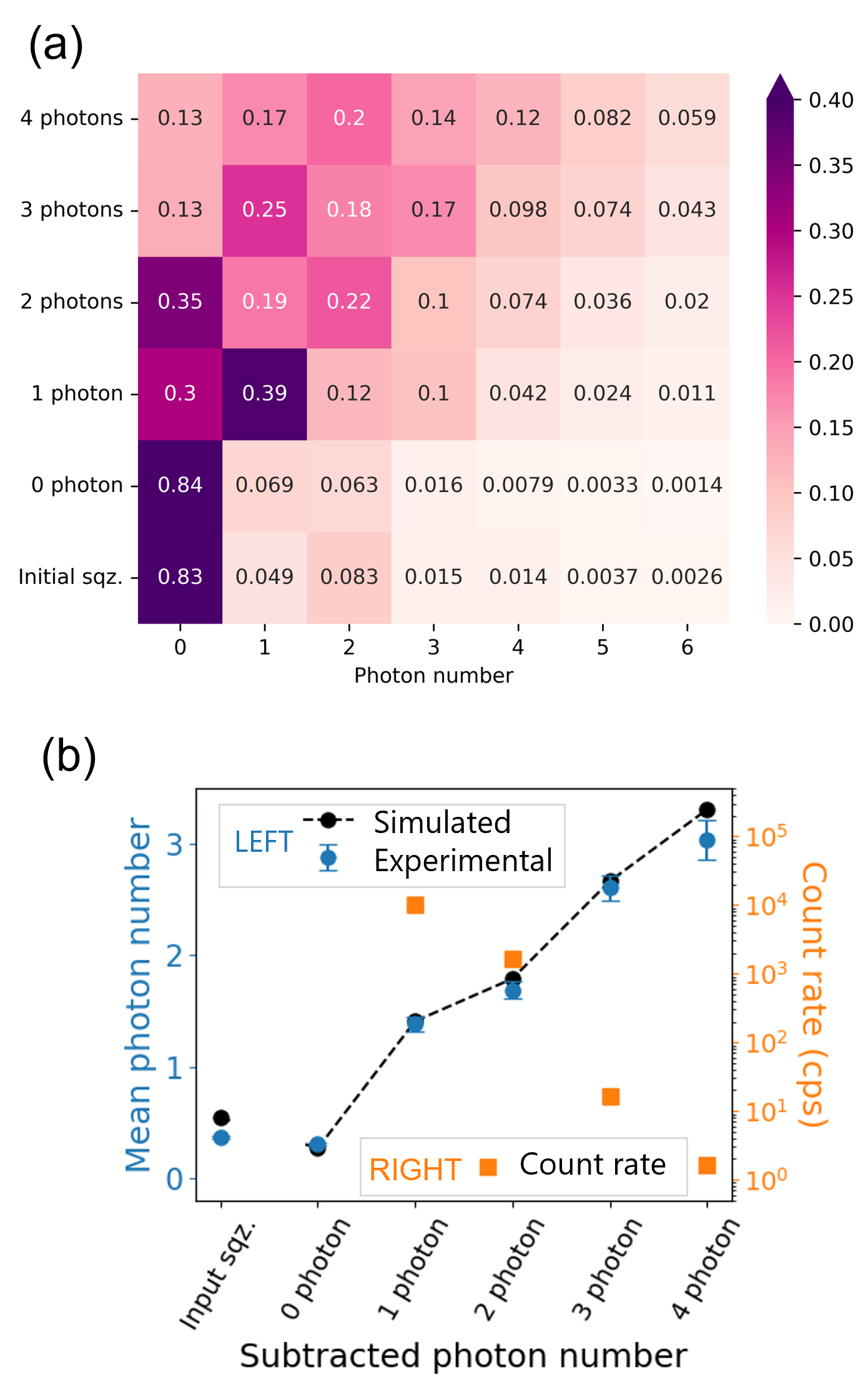}
\caption{(a) Experimentally obtained photon number distributions. (b) Mean photon numbers (left) and count rates (right). Blue circles with error bars: experimental results. Black circles: simulation results. Orange rectangles: count rates.}\label{photondist}
\end{figure}

Figure~\ref{photondist} (a)  shows heatmaps derived from the photon number distributions $\rho_{n,n}$ for each state. In the absence of loss, the input squeezed vacuum state consists solely of even-photon-number components. However, due to experimental losses, odd-photon-number components are also present.
Next, when one photon is subtracted, the photon number distribution shifts. This indicates that applying the annihilation operator shifts the photon number distribution by one $\hat{a}\ket{n}=\sqrt{n}\ket{n-1}\ (n>0)$. Moreover, the original zero-photon component is truncated $\hat{a}\ket{0}=0$, causing the both maximum and mean photon number of the cat state to increase as more photons are subtracted. This behavior is the same for cases where two or more photons are subtracted. 

Figure~\ref{photondist} (b) plots the simulated (black circles with dashed lines) and experimental (blue circles with error bars) mean photon numbers (on the left axis). The error bars for the experimental data represent the standard deviation ($1\sigma$) calculated using the bootstrap method. As can be seen from the results, the experimental data agrees very well with the simulation. 

Figure~\ref{photondist} (b) also plots generation rates on the right axis with orange rectangles. As evident from this figure, the generation rate decreases exponentially with the increase in photon number. In previous studies, the event rate for three-photon subtraction ranged from a few events per minute to a few events per second, rendering four-photon subtraction experiments impractical. In this study, leveraging a broadband squeezed light source, a high-speed photon number-resolving detector, and other advanced experimental techniques; we achieved an event rate of \SI{200}{cps} for three-photon and \SI{1.5}{cps} even for four-photon subtraction.

\subsection{Density matrix representation in the position and momentum bases}
\begin{figure}
\centering
\includegraphics[width=\textwidth]{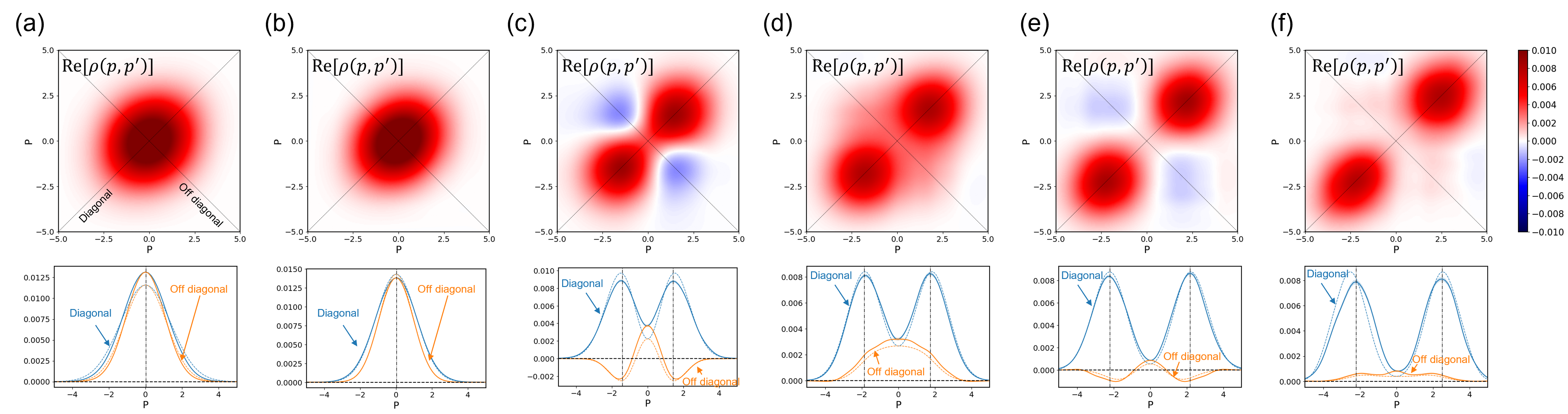}
\caption{(a-f) top: Heatmaps of the real part of density matrices in momentum bases $\mathrm{Re}[\rho(p,p')]$, bottom: diagonal and off-diagonal components (blue: $\mathrm{Re}[\rho(p,p)]$, orange: $\mathrm{Re}[\rho(p,-p)]$, respectively). The vertical dashed-dotted lines indicate the peak position of the diagonal components. The dashed lines in the bottom figures represent the simulation results. (a) Input squeezed vacuum and (b-f) zero- to four-photon subtraction, respectively.}\label{rho_xx_pp}
\end{figure}

The density matrices in the quadrature basis $\rho(p,p')$ were obtained from those in the photon number basis $\rho_{n,m}$ by 
\begin{align}
\rho(p,p')=\sum_{n,m}\rho_{n,m}\braket{p}{n}\braket{m}{p'}.
\end{align}
For each state, the real part of the density matrix in the momentum representation $\mathrm{Re}[\rho(p,p')]$ is shown as a heatmap in Fig.~\ref{rho_xx_pp}. Below each heatmap, the diagonal (blue) and off-diagonal (orange) components are plotted with error bars ($\mathrm{Re}[\rho(p,p)]$ and $\mathrm{Re}[\rho(p,-p)]$, respectively). The dashed-dotted lines indicate the peak positions of the diagonal components. The simulation results are also plotted in the lower plot as dashed lines of the same color. 
In Fig.~\ref{rho_xx_pp}(a) the input squeezed vacuum state and Fig.~\ref{rho_xx_pp}(b) the zero-photon subtraction case, the diagonal components exhibit only a single peak. Additionally, the off-diagonal components are narrower compared to the diagonal ones, which can be attributed to the effects of loss. 
Figure~\ref{rho_xx_pp}(c)–(f) show the results for one- to four-photon subtraction, respectively. In these cases, the diagonal components exhibit two peaks, corresponding to the ``alive cat'' and ``dead cat'', as explained in Fig.~\ref{fig:catstates} and Fig.~\ref{fig:photonsubtraction}. Significant off-diagonal components are also observed at the positions of these peaks, with their signs varying according to the number of subtracted photons. This indicates quantum interference between the two peaks, clearly distinguishing the states as cat states rather than mere mixtures of coherent states. It can be observed that the experimental results for the cat states with an increasing amplitude by the subtractions generally agree with the simulation results. The larger discrepancy in the case of single-photon subtraction is believed to be due to electrical noise in the TES signal, which was incorrectly detected as single-photon detection event. This issue can be improved by performing waveform analysis of the TES signal. 

The imaginary part of the density matrix and the position representation of the density matrix, not shown here, are included in the Supplementary Information \cite{supplement}.

\subsection{Wigner functions}
\begin{figure}
\centering
\includegraphics[width=\textwidth]{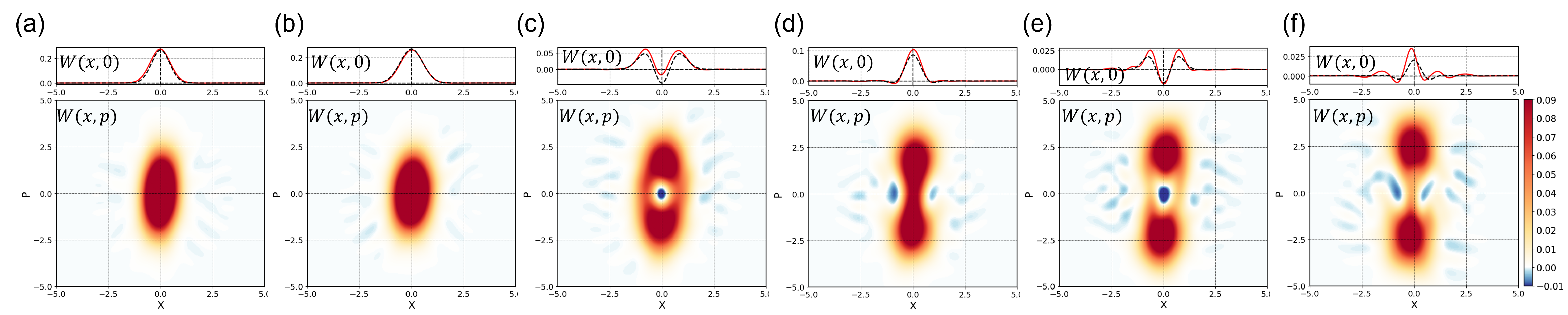}
\caption{Reconstructed Wigner functions without loss correction. The plots shown above the heatmaps represent the cross-section at $p=0$ with error bars (red). The simulation results are also shown in black-dashed lines.  (a) Input squeezed vacuum and (b-f) zero- to four-photon subtraction, respectively.}\label{wigners}
\end{figure}

The Wigner functions for each state are presented in Fig.~\ref{wigners}. A key feature is the presence of negative values in the Wigner functions for all states after photon subtraction (c)-(f). The sign of the Wigner function at the origin indicates the parity of the state: positive for even $n$ and negative for odd $n$. Remarkably, the negative values in the Wigner function persist even after four-photon subtractions, demonstrating the action of the annihilation operator four times without the need for loss correction. 

\section{Conclusions}\label{conclusions}
In this work, we successfully demonstrated the high-rate photon subtraction of up to four photons from a squeezed vacuum state using a broadband OPA and the high-efficiency TES. Despite the system's sensitivity to residual optical losses and phase instability, the experimental results showed excellent agreement with simulations, confirming the non-classical nature of the generated states through Wigner function negativity, and quantum coherence observed in the density matrix in the momentum representation for the growing cat state without any data correction.

This achievement represents a significant breakthrough in generating high-amplitude cat states, which serve as essential resource states for creating logical qubits indispensable for fault-tolerant quantum computing \cite{Aghaee2025}. The scalability of the experimental setup enables future extensions to photon addition \cite{Parigi2007,Chen2024, Fadrny2024}, generalized photon subtraction \cite{Takase2021}, quantum state synthesizer \cite{Su2019}, and GKP qubit synthesizer \cite{Takase2023}. These techniques can be further developed for applications such as the high-quality and high-rate generation of Schr\"{o}dinger cat states and GKP qubits, which are essential for fault-tolerant optical quantum computers.

By advancing the generation of complex quantum states, this work paves the way for new applications in quantum sensing, quantum communication, as well as high-speed optical quantum computing.

\section{Methods}\label{method}
\subsection{Details of WG OPA}
We use a periodically poled lithium niobate (PPLN) waveguide as the OPA for squeezed light generation. The core of the waveguide consists of ZnO-doped LN and is directly bonded to a lithium tantalate substrate. The periodic poling structure for quasi-phase matching (QPM) is fabricated using an electrical poling method \cite{Umeki2010}. The poling cycle is approximately \SI{18}{\micro m}, and the structure is designed as a type-0 PPLN. The waveguide is \SI{10}{mm} long, and its core has a width of about \SI{9}{\micro m} and a thickness of about \SI{8}{\micro m}. The waveguide structure is fabricated using a mechanical dicing saw. More details on the fabrication can be found in the reference \cite{Kashiwazaki2021}. Both the input and output end faces of the waveguide are coated with anti-reflective coatings for both 1545.32-nm and 772.66-nm light. The temperature of the waveguide is controlled to maintain the QPM condition.

\subsection{Details of Ti-Au TES}
The TES device was fabricated at the superconducting quantum circuit fabrication facility (Qufab) at AIST. The superconducting TES film comprises bilayers of titanium and gold, deposited by DC magnetron sputtering to thicknesses of \SI{20}{nm} and \SI{10}{nm}, respectively. To enhance detection efficiency, an optical cavity structure was adopted, in which the TES films are embedded between a high-reflection mirror and anti-reflection coatings. The TES is patterned into a square with dimensions of \SI{5}{\micro m} $\times$ \SI{5}{\micro m}. Superconducting niobium leads were fabricated on two opposite sides of the TES, and the resulting resistance changes are read via these leads using a SQUID-based current amplifier.

To deliver photons, a high-NA optical fiber (UHNA7, Nufern) with a mode field diameter (MFD) of \SI{3.2}{\micro m} was coupled to the TES device. The fiber-coupled TES module was installed in an adiabatic demagnetization refrigerator and cooled to below \SI{280}{mK}. Inside the refrigerator, the high-NA optical fiber was spliced to an SMF28 fiber (Cornig) to convert the MFD to \SI{10}{\micro m}, with an MFD conversion loss of less than \SI{0.2}{dB}.

The critical temperature and normal resistance of the TES are \SI{308.6}{mK} and \SI{2.653}{\ohm}, respectively. The TES exhibits an energy resolution of $\Delta E=\SI{0.176}{eV}$, corresponding to an $E/\Delta E$ ratio of 4.54 for a single-photon energy of \SI{0.8}{eV} at a wavelength of \SI{1.5}{\micro m}, thereby demonstrating sufficient capability to resolve individual photon states. The system detection efficiency of the TES was carefully calibrated using an evaluation system traceable to the National Metrology Institute of Japan (NMIJ) at AIST, yielding an efficiency of 89.2\% at a wavelength of \SI{1.5}{\micro m}. The decay time constant of the observed signals was \SI{107}{ns}, enabling a repetition frequency of up to \SI{5}{MHz}.

\backmatter

\bmhead{Supplementary information}
This submission is accompanied by Supplementary Information, which includes detailed experimental apparatus and additional data.

\bmhead{Acknowledgements}
This work was partly supported by Japan Science and Technology (JST) Agency (Moonshot R\&D, Grant No. JPMJMS2064, and PRESTO, Grant No. JPMJPR2254), the UTokyo Foundation, and donations from Nichia Corporation. W.A. acknowledges the funding from Japan Society for the Promotion of Science (JSPS) KAKENHI (Grant No. 23K13040). M.E., K.T, and W.A. acknowledge supports from the Research Foundation for Opto-Science and Technology. P.M. acknowledges the project 25-17472S of the Czech Science Foundation, the European Union’s HORIZON Research and Innovation Actions under Grant Agreement no. 101080173 (CLUSTEC) and the project CZ.02.01.01\/00\/22\_008\/0004649 (QUEENTEC) of EU and the Czech Ministry of Education, Youth and Sport. R.F. acknowledges the project No. 21-13265X of the Czech Science Foundation.

\bmhead{Author contributions}
M.E. conceived the project and led the experiment. M.E. built the optical and electrical setup with support from T.N., T.S., K. Takahashi, and R.N.. M.E. coded the control program and the FPGA program for data acquisition with support from T.N.. M.E., P.M., and R.F. performed simulations and theory evaluation, and analyzed the data. T.K., A.I., and T.U. provided the OPA used in the experiment. S.T. and D.F. provided the TES used in the experiment. M.E. wrote the manuscript with P.M., R.F., K. Takase, W.A., A.F., and all the co-authors.

\bmhead{Competing interests}
The authors declare no competing interests.

\bmhead{Data and materials availability}
All data are available either in the manuscript or in the supplementary information.



\begin{thebibliography}{42}
\ifx \bisbn   \undefined \def \bisbn  #1{ISBN #1}\fi
\ifx \binits  \undefined \def \binits#1{#1}\fi
\ifx \bauthor  \undefined \def \bauthor#1{#1}\fi
\ifx \batitle  \undefined \def \batitle#1{#1}\fi
\ifx \bjtitle  \undefined \def \bjtitle#1{#1}\fi
\ifx \bvolume  \undefined \def \bvolume#1{\textbf{#1}}\fi
\ifx \byear  \undefined \def \byear#1{#1}\fi
\ifx \bissue  \undefined \def \bissue#1{#1}\fi
\ifx \bfpage  \undefined \def \bfpage#1{#1}\fi
\ifx \blpage  \undefined \def \blpage #1{#1}\fi
\ifx \burl  \undefined \def \burl#1{\textsf{#1}}\fi
\ifx \doiurl  \undefined \def \doiurl#1{\url{https://doi.org/#1}}\fi
\ifx \betal  \undefined \def \betal{\textit{et al.}}\fi
\ifx \binstitute  \undefined \def \binstitute#1{#1}\fi
\ifx \binstitutionaled  \undefined \def \binstitutionaled#1{#1}\fi
\ifx \bctitle  \undefined \def \bctitle#1{#1}\fi
\ifx \beditor  \undefined \def \beditor#1{#1}\fi
\ifx \bpublisher  \undefined \def \bpublisher#1{#1}\fi
\ifx \bbtitle  \undefined \def \bbtitle#1{#1}\fi
\ifx \bedition  \undefined \def \bedition#1{#1}\fi
\ifx \bseriesno  \undefined \def \bseriesno#1{#1}\fi
\ifx \blocation  \undefined \def \blocation#1{#1}\fi
\ifx \bsertitle  \undefined \def \bsertitle#1{#1}\fi
\ifx \bsnm \undefined \def \bsnm#1{#1}\fi
\ifx \bsuffix \undefined \def \bsuffix#1{#1}\fi
\ifx \bparticle \undefined \def \bparticle#1{#1}\fi
\ifx \barticle \undefined \def \barticle#1{#1}\fi
\bibcommenthead
\ifx \bconfdate \undefined \def \bconfdate #1{#1}\fi
\ifx \botherref \undefined \def \botherref #1{#1}\fi
\ifx \url \undefined \def \url#1{\textsf{#1}}\fi
\ifx \bchapter \undefined \def \bchapter#1{#1}\fi
\ifx \bbook \undefined \def \bbook#1{#1}\fi
\ifx \bcomment \undefined \def \bcomment#1{#1}\fi
\ifx \oauthor \undefined \def \oauthor#1{#1}\fi
\ifx \citeauthoryear \undefined \def \citeauthoryear#1{#1}\fi
\ifx \endbibitem  \undefined \def \endbibitem {}\fi
\ifx \bconflocation  \undefined \def \bconflocation#1{#1}\fi
\ifx \arxivurl  \undefined \def \arxivurl#1{\textsf{#1}}\fi
\csname PreBibitemsHook\endcsname

\bibitem[\protect\citeauthoryear{Lloyd and Braunstein}{1999}]{Lloyd1999}
\begin{barticle}
\bauthor{\bsnm{Lloyd}, \binits{S.}},
\bauthor{\bsnm{Braunstein}, \binits{S.L.}}:
\batitle{Quantum computation over continuous variables}.
\bjtitle{Physical Review Letters}
\bvolume{82}(\bissue{8}),
\bfpage{1784}--\blpage{1787}
(\byear{1999})
\doiurl{10.1103/PhysRevLett.82.1784}
\end{barticle}
\endbibitem

\bibitem[\protect\citeauthoryear{Takeda and Furusawa}{2019}]{Takeda2019}
\begin{barticle}
\bauthor{\bsnm{Takeda}, \binits{S.}},
\bauthor{\bsnm{Furusawa}, \binits{A.}}:
\batitle{Toward large-scale fault-tolerant universal photonic quantum computing}.
\bjtitle{APL Photonics}
\bvolume{4}(\bissue{6}),
\bfpage{060902}
(\byear{2019})
\doiurl{10.1063/1.5100160}
\end{barticle}
\endbibitem

\bibitem[\protect\citeauthoryear{Asavanant et~al.}{2019}]{Asavanant2019}
\begin{barticle}
\bauthor{\bsnm{Asavanant}, \binits{W.}},
\bauthor{\bsnm{Shiozawa}, \binits{Y.}},
\bauthor{\bsnm{Yokoyama}, \binits{S.}},
\bauthor{\bsnm{Charoensombutamon}, \binits{B.}},
\bauthor{\bsnm{Emura}, \binits{H.}},
\bauthor{\bsnm{Alexander}, \binits{R.N.}},
\bauthor{\bsnm{Takeda}, \binits{S.}},
\bauthor{\bsnm{Yoshikawa}, \binits{J.I.}},
\bauthor{\bsnm{Menicucci}, \binits{N.C.}},
\bauthor{\bsnm{Yonezawa}, \binits{H.}},
\bauthor{\bsnm{Furusawa}, \binits{A.}}:
\batitle{Generation of time-domain-multiplexed two-dimensional cluster state}.
\bjtitle{Science}
\bvolume{366}(\bissue{6463}),
\bfpage{373}--\blpage{376}
(\byear{2019})
\doiurl{10.1126/science.aay2645}
\end{barticle}
\endbibitem

\bibitem[\protect\citeauthoryear{Larsen et~al.}{2019}]{Larsen2019}
\begin{barticle}
\bauthor{\bsnm{Larsen}, \binits{M.V.}},
\bauthor{\bsnm{Guo}, \binits{X.}},
\bauthor{\bsnm{Breum}, \binits{C.R.}},
\bauthor{\bsnm{Neergaard-Nielsen}, \binits{J.S.}},
\bauthor{\bsnm{Andersen}, \binits{U.L.}}:
\batitle{Deterministic generation of a two-dimensional cluster state}.
\bjtitle{Science}
\bvolume{366}(\bissue{6463}),
\bfpage{369}--\blpage{372}
(\byear{2019})
\doiurl{10.1126/science.aay4354}
\end{barticle}
\endbibitem

\bibitem[\protect\citeauthoryear{Aghaee~Rad et~al.}{2025}]{Aghaee2025}
\begin{barticle}
\bauthor{\bsnm{Aghaee~Rad}, \binits{H.}},
\bauthor{\bsnm{Ainsworth}, \binits{T.}},
\bauthor{\bsnm{Alexander}, \binits{R.N.}},
\bauthor{\bsnm{Altieri}, \binits{B.}},
\bauthor{\bsnm{Askarani}, \binits{M.F.}},
\bauthor{\bsnm{Baby}, \binits{R.}},
\bauthor{\bsnm{Banchi}, \binits{L.}},
\bauthor{\bsnm{Baragiola}, \binits{B.Q.}},
\bauthor{\bsnm{Bourassa}, \binits{J.E.}},
\bauthor{\bsnm{Chadwick}, \binits{R.S.}},
\bauthor{\bsnm{Charania}, \binits{I.}},
\bauthor{\bsnm{Chen}, \binits{H.}},
\bauthor{\bsnm{Collins}, \binits{M.J.}},
\bauthor{\bsnm{Contu}, \binits{P.}},
\bauthor{\bsnm{D’Arcy}, \binits{N.}},
\bauthor{\bsnm{Dauphinais}, \binits{G.}},
\bauthor{\bsnm{De~Prins}, \binits{R.}},
\bauthor{\bsnm{Deschenes}, \binits{D.}},
\bauthor{\bsnm{Di~Luch}, \binits{I.}},
\bauthor{\bsnm{Duque}, \binits{S.}},
\bauthor{\bsnm{Edke}, \binits{P.}},
\bauthor{\bsnm{Fayer}, \binits{S.E.}},
\bauthor{\bsnm{Ferracin}, \binits{S.}},
\bauthor{\bsnm{Ferretti}, \binits{H.}},
\bauthor{\bsnm{Gefaell}, \binits{J.}},
\bauthor{\bsnm{Glancy}, \binits{S.}},
\bauthor{\bsnm{González-Arciniegas}, \binits{C.}},
\bauthor{\bsnm{Grainge}, \binits{T.}},
\bauthor{\bsnm{Han}, \binits{Z.}},
\bauthor{\bsnm{Hastrup}, \binits{J.}},
\bauthor{\bsnm{Helt}, \binits{L.G.}},
\bauthor{\bsnm{Hillmann}, \binits{T.}},
\bauthor{\bsnm{Hundal}, \binits{J.}},
\bauthor{\bsnm{Izumi}, \binits{S.}},
\bauthor{\bsnm{Jaeken}, \binits{T.}},
\bauthor{\bsnm{Jonas}, \binits{M.}},
\bauthor{\bsnm{Kocsis}, \binits{S.}},
\bauthor{\bsnm{Krasnokutska}, \binits{I.}},
\bauthor{\bsnm{Larsen}, \binits{M.V.}},
\bauthor{\bsnm{Laskowski}, \binits{P.}},
\bauthor{\bsnm{Laudenbach}, \binits{F.}},
\bauthor{\bsnm{Lavoie}, \binits{J.}},
\bauthor{\bsnm{Li}, \binits{M.}},
\bauthor{\bsnm{Lomonte}, \binits{E.}},
\bauthor{\bsnm{Lopetegui}, \binits{C.E.}},
\bauthor{\bsnm{Luey}, \binits{B.}},
\bauthor{\bsnm{Lund}, \binits{A.P.}},
\bauthor{\bsnm{Ma}, \binits{C.}},
\bauthor{\bsnm{Madsen}, \binits{L.S.}},
\bauthor{\bsnm{Mahler}, \binits{D.H.}},
\bauthor{\bsnm{Mantilla~Calderón}, \binits{L.}},
\bauthor{\bsnm{Menotti}, \binits{M.}},
\bauthor{\bsnm{Miatto}, \binits{F.M.}},
\bauthor{\bsnm{Morrison}, \binits{B.}},
\bauthor{\bsnm{Nadkarni}, \binits{P.J.}},
\bauthor{\bsnm{Nakamura}, \binits{T.}},
\bauthor{\bsnm{Neuhaus}, \binits{L.}},
\bauthor{\bsnm{Niu}, \binits{Z.}},
\bauthor{\bsnm{Noro}, \binits{R.}},
\bauthor{\bsnm{Papirov}, \binits{K.}},
\bauthor{\bsnm{Pesah}, \binits{A.}},
\bauthor{\bsnm{Phillips}, \binits{D.S.}},
\bauthor{\bsnm{Plick}, \binits{W.N.}},
\bauthor{\bsnm{Rogalsky}, \binits{T.}},
\bauthor{\bsnm{Rortais}, \binits{F.}},
\bauthor{\bsnm{Sabines-Chesterking}, \binits{J.}},
\bauthor{\bsnm{Safavi-Bayat}, \binits{S.}},
\bauthor{\bsnm{Sazhaev}, \binits{E.}},
\bauthor{\bsnm{Seymour}, \binits{M.}},
\bauthor{\bsnm{Rezaei~Shad}, \binits{K.}},
\bauthor{\bsnm{Silverman}, \binits{M.}},
\bauthor{\bsnm{Srinivasan}, \binits{S.A.}},
\bauthor{\bsnm{Stephan}, \binits{M.}},
\bauthor{\bsnm{Tang}, \binits{Q.Y.}},
\bauthor{\bsnm{Tasker}, \binits{J.F.}},
\bauthor{\bsnm{Teo}, \binits{Y.S.}},
\bauthor{\bsnm{Then}, \binits{R.B.}},
\bauthor{\bsnm{Tremblay}, \binits{J.E.}},
\bauthor{\bsnm{Tzitrin}, \binits{I.}},
\bauthor{\bsnm{Vaidya}, \binits{V.D.}},
\bauthor{\bsnm{Vasmer}, \binits{M.}},
\bauthor{\bsnm{Vernon}, \binits{Z.}},
\bauthor{\bsnm{Villalobos}, \binits{L.F.S.S.M.}},
\bauthor{\bsnm{Walshe}, \binits{B.W.}},
\bauthor{\bsnm{Weil}, \binits{R.}},
\bauthor{\bsnm{Xin}, \binits{X.}},
\bauthor{\bsnm{Yan}, \binits{X.}},
\bauthor{\bsnm{Yao}, \binits{Y.}},
\bauthor{\bsnm{Zamani~Abnili}, \binits{M.}},
\bauthor{\bsnm{Zhang}, \binits{Y.}}:
\batitle{Scaling and networking a modular photonic quantum computer}.
\bjtitle{Nature}
(\byear{2025})
\doiurl{10.1038/s41586-024-08406-9}
\end{barticle}
\endbibitem

\bibitem[\protect\citeauthoryear{Sudbeck et~al.}{2020}]{Sudbeck2020}
\begin{barticle}
\bauthor{\bsnm{Sudbeck}, \binits{J.}},
\bauthor{\bsnm{Steinlechner}, \binits{S.}},
\bauthor{\bsnm{Korobko}, \binits{M.}},
\bauthor{\bsnm{Schnabel}, \binits{R.}}:
\batitle{Demonstration of interferometer enhancement through {EPR} entanglement}.
\bjtitle{Nature Photonics}
\bvolume{14}(\bissue{4}),
\bfpage{240}--\blpage{244}
(\byear{2020})
\doiurl{10.1038/s41566-019-0583-3}
\end{barticle}
\endbibitem

\bibitem[\protect\citeauthoryear{Anisimov et~al.}{2010}]{Anisimov2010}
\begin{barticle}
\bauthor{\bsnm{Anisimov}, \binits{P.M.}},
\bauthor{\bsnm{Raterman}, \binits{G.M.}},
\bauthor{\bsnm{Chiruvelli}, \binits{A.}},
\bauthor{\bsnm{Plick}, \binits{W.N.}},
\bauthor{\bsnm{Huver}, \binits{S.D.}},
\bauthor{\bsnm{Lee}, \binits{H.}},
\bauthor{\bsnm{Dowling}, \binits{J.P.}}:
\batitle{Quantum metrology with two-mode squeezed vacuum: parity detection beats the heisenberg limit}.
\bjtitle{Physical Review Letters}
\bvolume{104}(\bissue{10}),
\bfpage{103602}
(\byear{2010})
\doiurl{10.1103/PhysRevLett.104.103602}
\end{barticle}
\endbibitem

\bibitem[\protect\citeauthoryear{Couteau et~al.}{2023}]{Couteau2023}
\begin{barticle}
\bauthor{\bsnm{Couteau}, \binits{C.}},
\bauthor{\bsnm{Barz}, \binits{S.}},
\bauthor{\bsnm{Durt}, \binits{T.}},
\bauthor{\bsnm{Gerrits}, \binits{T.}},
\bauthor{\bsnm{Huwer}, \binits{J.}},
\bauthor{\bsnm{Prevedel}, \binits{R.}},
\bauthor{\bsnm{Rarity}, \binits{J.}},
\bauthor{\bsnm{Shields}, \binits{A.}},
\bauthor{\bsnm{Weihs}, \binits{G.}}:
\batitle{Applications of single photons to quantum communication and computing}.
\bjtitle{Nature Reviews Physics}
\bvolume{5}(\bissue{6}),
\bfpage{326}--\blpage{338}
(\byear{2023})
\doiurl{10.1038/s42254-023-00583-2}
\end{barticle}
\endbibitem

\bibitem[\protect\citeauthoryear{Gottesman et~al.}{2001}]{Gottesman2001}
\begin{barticle}
\bauthor{\bsnm{Gottesman}, \binits{D.}},
\bauthor{\bsnm{Kitaev}, \binits{A.}},
\bauthor{\bsnm{Preskill}, \binits{J.}}:
\batitle{Encoding a qubit in an oscillator}.
\bjtitle{Physical Review A}
\bvolume{64}(\bissue{1}),
\bfpage{012310}
(\byear{2001})
\doiurl{10.1103/PhysRevA.64.012310}
\end{barticle}
\endbibitem

\bibitem[\protect\citeauthoryear{Baragiola et~al.}{2019}]{Baragiola2019}
\begin{barticle}
\bauthor{\bsnm{Baragiola}, \binits{B.Q.}},
\bauthor{\bsnm{Pantaleoni}, \binits{G.}},
\bauthor{\bsnm{Alexander}, \binits{R.N.}},
\bauthor{\bsnm{Karanjai}, \binits{A.}},
\bauthor{\bsnm{Menicucci}, \binits{N.C.}}:
\batitle{All-gaussian universality and fault tolerance with the {G}ottesman-{K}itaev-{P}reskill code}.
\bjtitle{Physical Review Letters}
\bvolume{123}(\bissue{20}),
\bfpage{200502}
(\byear{2019})
\doiurl{10.1103/PhysRevLett.123.200502}
\end{barticle}
\endbibitem

\bibitem[\protect\citeauthoryear{Konno et~al.}{2021}]{Konno2021}
\begin{botherref}
\oauthor{\bsnm{Konno}, \binits{S.}},
\oauthor{\bsnm{Asavanant}, \binits{W.}},
\oauthor{\bsnm{Fukui}, \binits{K.}},
\oauthor{\bsnm{Sakaguchi}, \binits{A.}},
\oauthor{\bsnm{Hanamura}, \binits{F.}},
\oauthor{\bsnm{Marek}, \binits{P.}},
\oauthor{\bsnm{Filip}, \binits{R.}},
\oauthor{\bsnm{Yoshikawa}, \binits{J.}},
\oauthor{\bsnm{Furusawa}, \binits{A.}}:
Non-clifford gate on optical qubits by nonlinear feedforward.
Physical Review Research
\textbf{3}(4)
(2021)
\doiurl{10.1103/PhysRevResearch.3.043026}
\end{botherref}
\endbibitem

\bibitem[\protect\citeauthoryear{Vasconcelos et~al.}{2010}]{Vasconcelos2010}
\begin{barticle}
\bauthor{\bsnm{Vasconcelos}, \binits{H.M.}},
\bauthor{\bsnm{Sanz}, \binits{L.}},
\bauthor{\bsnm{Glancy}, \binits{S.}}:
\batitle{All-optical generation of states for "encoding a qubit in an oscillator"}.
\bjtitle{Optics Letters}
\bvolume{35}(\bissue{19}),
\bfpage{3261}--\blpage{3}
(\byear{2010})
\doiurl{10.1364/OL.35.003261}
\end{barticle}
\endbibitem

\bibitem[\protect\citeauthoryear{Weigand and Terhal}{2018}]{Weigand2018}
\begin{barticle}
\bauthor{\bsnm{Weigand}, \binits{D.J.}},
\bauthor{\bsnm{Terhal}, \binits{B.M.}}:
\batitle{Generating grid states from {S}chrödinger-cat states without postselection}.
\bjtitle{Physical Review A}
\bvolume{97}(\bissue{2}),
\bfpage{022341}
(\byear{2018})
\doiurl{10.1103/PhysRevA.97.022341}
\end{barticle}
\endbibitem

\bibitem[\protect\citeauthoryear{Eaton et~al.}{2019}]{Eaton2019}
\begin{botherref}
\oauthor{\bsnm{Eaton}, \binits{M.}},
\oauthor{\bsnm{Nehra}, \binits{R.}},
\oauthor{\bsnm{Pfister}, \binits{O.}}:
Non-gaussian and gottesman–kitaev–preskill state preparation by photon catalysis.
New Journal of Physics
\textbf{21}(11)
(2019)
\doiurl{10.1088/1367-2630/ab5330}
\end{botherref}
\endbibitem

\bibitem[\protect\citeauthoryear{Tzitrin et~al.}{2020}]{Tzitrin2020}
\begin{botherref}
\oauthor{\bsnm{Tzitrin}, \binits{I.}},
\oauthor{\bsnm{Bourassa}, \binits{J.E.}},
\oauthor{\bsnm{Menicucci}, \binits{N.C.}},
\oauthor{\bsnm{Sabapathy}, \binits{K.K.}}:
Progress towards practical qubit computation using approximate {G}ottesman-{K}itaev-{P}reskill codes.
Physical Review A
\textbf{101}(3)
(2020)
\doiurl{10.1103/PhysRevA.101.032315}
\end{botherref}
\endbibitem

\bibitem[\protect\citeauthoryear{Takase et~al.}{2023}]{Takase2023}
\begin{botherref}
\oauthor{\bsnm{Takase}, \binits{K.}},
\oauthor{\bsnm{Fukui}, \binits{K.}},
\oauthor{\bsnm{Kawasaki}, \binits{A.}},
\oauthor{\bsnm{Asavanant}, \binits{W.}},
\oauthor{\bsnm{Endo}, \binits{M.}},
\oauthor{\bsnm{Yoshikawa}, \binits{J.}},
\oauthor{\bsnm{Loock}, \binits{P.}},
\oauthor{\bsnm{Furusawa}, \binits{A.}}:
{G}ottesman-{K}itaev-{P}reskill qubit synthesizer for propagating light.
npj Quantum Information
\textbf{9}(1)
(2023)
\doiurl{10.1038/s41534-023-00772-y}
\end{botherref}
\endbibitem

\bibitem[\protect\citeauthoryear{Cochrane et~al.}{1999}]{Cochrane1999}
\begin{barticle}
\bauthor{\bsnm{Cochrane}, \binits{P.T.}},
\bauthor{\bsnm{Milburn}, \binits{G.J.}},
\bauthor{\bsnm{Munro}, \binits{W.J.}}:
\batitle{Macroscopically distinct quantum-superposition states as a bosonic code for amplitude damping}.
\bjtitle{Physical Review A}
\bvolume{59}(\bissue{4}),
\bfpage{2631}--\blpage{2634}
(\byear{1999})
\doiurl{10.1103/PhysRevA.59.2631}
\end{barticle}
\endbibitem

\bibitem[\protect\citeauthoryear{Ralph et~al.}{2003}]{Ralph2003}
\begin{botherref}
\oauthor{\bsnm{Ralph}, \binits{T.C.}},
\oauthor{\bsnm{Gilchrist}, \binits{A.}},
\oauthor{\bsnm{Milburn}, \binits{G.J.}},
\oauthor{\bsnm{Munro}, \binits{W.J.}},
\oauthor{\bsnm{Glancy}, \binits{S.}}:
Quantum computation with optical coherent states.
Physical Review A
\textbf{68}(4)
(2003)
\doiurl{10.1103/PhysRevA.68.042319}
\end{botherref}
\endbibitem

\bibitem[\protect\citeauthoryear{Konno et~al.}{2024}]{Konno2024}
\begin{barticle}
\bauthor{\bsnm{Konno}, \binits{S.}},
\bauthor{\bsnm{Asavanant}, \binits{W.}},
\bauthor{\bsnm{Hanamura}, \binits{F.}},
\bauthor{\bsnm{Nagayoshi}, \binits{H.}},
\bauthor{\bsnm{Fukui}, \binits{K.}},
\bauthor{\bsnm{Sakaguchi}, \binits{A.}},
\bauthor{\bsnm{Ide}, \binits{R.}},
\bauthor{\bsnm{China}, \binits{F.}},
\bauthor{\bsnm{Yabuno}, \binits{M.}},
\bauthor{\bsnm{Miki}, \binits{S.}},
\bauthor{\bsnm{Terai}, \binits{H.}},
\bauthor{\bsnm{Takase}, \binits{K.}},
\bauthor{\bsnm{Endo}, \binits{M.}},
\bauthor{\bsnm{Marek}, \binits{P.}},
\bauthor{\bsnm{Filip}, \binits{R.}},
\bauthor{\bsnm{Loock}, \binits{P.}},
\bauthor{\bsnm{Furusawa}, \binits{A.}}:
\batitle{Logical states for fault-tolerant quantum computation with propagating light}.
\bjtitle{Science}
\bvolume{383}(\bissue{6680}),
\bfpage{289}--\blpage{293}
(\byear{2024})
\doiurl{10.1126/science.adk7560}
\end{barticle}
\endbibitem

\bibitem[\protect\citeauthoryear{Dakna et~al.}{1997}]{Dakna1997}
\begin{barticle}
\bauthor{\bsnm{Dakna}, \binits{M.}},
\bauthor{\bsnm{Anhut}, \binits{T.}},
\bauthor{\bsnm{Opatrný}, \binits{T.}},
\bauthor{\bsnm{Knöll}, \binits{L.}},
\bauthor{\bsnm{Welsch}, \binits{D.G.}}:
\batitle{Generating {S}chrödinger-cat-like states by means of conditional measurements on a beam splitter}.
\bjtitle{Physical Review A}
\bvolume{55}(\bissue{4}),
\bfpage{3184}--\blpage{3194}
(\byear{1997})
\doiurl{10.1103/PhysRevA.55.3184}
\end{barticle}
\endbibitem

\bibitem[\protect\citeauthoryear{Ourjoumtsev et~al.}{2006}]{Ourjoumtsev2006}
\begin{barticle}
\bauthor{\bsnm{Ourjoumtsev}, \binits{A.}},
\bauthor{\bsnm{Tualle-Brouri}, \binits{R.}},
\bauthor{\bsnm{Laurat}, \binits{J.}},
\bauthor{\bsnm{Grangier}, \binits{P.}}:
\batitle{Generating optical {S}chrödinger kittens for quantum information processing}.
\bjtitle{Science}
\bvolume{312}(\bissue{5770}),
\bfpage{83}--\blpage{6}
(\byear{2006})
\doiurl{10.1126/science.1122858}
\end{barticle}
\endbibitem

\bibitem[\protect\citeauthoryear{Neergaard-Nielsen}{2008}]{Neergaard-Nielsen2008}
\begin{botherref}
\oauthor{\bsnm{Neergaard-Nielsen}, \binits{J.S.}}:
Generation of single photons and {S}chrödinger kitten states of light.
Thesis,
University of Copenhagen
(2008)
\end{botherref}
\endbibitem

\bibitem[\protect\citeauthoryear{Asavanant et~al.}{2017}]{Asavanant2017}
\begin{barticle}
\bauthor{\bsnm{Asavanant}, \binits{W.}},
\bauthor{\bsnm{Nakashima}, \binits{K.}},
\bauthor{\bsnm{Shiozawa}, \binits{Y.}},
\bauthor{\bsnm{Yoshikawa}, \binits{J.}},
\bauthor{\bsnm{Furusawa}, \binits{A.}}:
\batitle{Generation of highly pure {S}chrödinger’s cat states and real-time quadrature measurements via optical filtering}.
\bjtitle{Optics Express}
\bvolume{25}(\bissue{26}),
\bfpage{32227}
(\byear{2017})
\doiurl{10.1364/oe.25.032227}
\end{barticle}
\endbibitem

\bibitem[\protect\citeauthoryear{Takase et~al.}{2021}]{Takase2021}
\begin{barticle}
\bauthor{\bsnm{Takase}, \binits{K.}},
\bauthor{\bsnm{Yoshikawa}, \binits{J.}},
\bauthor{\bsnm{Asavanant}, \binits{W.}},
\bauthor{\bsnm{Endo}, \binits{M.}},
\bauthor{\bsnm{Furusawa}, \binits{A.}}:
\batitle{Generation of optical {S}chrödinger cat states by generalized photon subtraction}.
\bjtitle{Physical Review A}
\bvolume{103}(\bissue{1}),
\bfpage{013710}
(\byear{2021})
\doiurl{10.1103/PhysRevA.103.013710}
\end{barticle}
\endbibitem

\bibitem[\protect\citeauthoryear{Gerrits et~al.}{2010}]{Gerrits2010}
\begin{barticle}
\bauthor{\bsnm{Gerrits}, \binits{T.}},
\bauthor{\bsnm{Glancy}, \binits{S.}},
\bauthor{\bsnm{Clement}, \binits{T.S.}},
\bauthor{\bsnm{Calkins}, \binits{B.}},
\bauthor{\bsnm{Lita}, \binits{A.E.}},
\bauthor{\bsnm{Miller}, \binits{A.J.}},
\bauthor{\bsnm{Migdall}, \binits{A.L.}},
\bauthor{\bsnm{Nam}, \binits{S.W.}},
\bauthor{\bsnm{Mirin}, \binits{R.P.}},
\bauthor{\bsnm{Knill}, \binits{E.}}:
\batitle{Generation of optical coherent-state superpositions by number-resolved photon subtraction from the squeezed vacuum}.
\bjtitle{Physical Review A}
\bvolume{82}(\bissue{3}),
\bfpage{031802}
(\byear{2010})
\doiurl{10.1103/PhysRevA.82.031802}
\end{barticle}
\endbibitem

\bibitem[\protect\citeauthoryear{Endo et~al.}{2023}]{Endo2023}
\begin{barticle}
\bauthor{\bsnm{Endo}, \binits{M.}},
\bauthor{\bsnm{He}, \binits{R.}},
\bauthor{\bsnm{Sonoyama}, \binits{T.}},
\bauthor{\bsnm{Takahashi}, \binits{K.}},
\bauthor{\bsnm{Kashiwazaki}, \binits{T.}},
\bauthor{\bsnm{Umeki}, \binits{T.}},
\bauthor{\bsnm{Takasu}, \binits{S.}},
\bauthor{\bsnm{Hattori}, \binits{K.}},
\bauthor{\bsnm{Fukuda}, \binits{D.}},
\bauthor{\bsnm{Fukui}, \binits{K.}},
\bauthor{\bsnm{Takase}, \binits{K.}},
\bauthor{\bsnm{Asavanant}, \binits{W.}},
\bauthor{\bsnm{Marek}, \binits{P.}},
\bauthor{\bsnm{Filip}, \binits{R.}},
\bauthor{\bsnm{Furusawa}, \binits{A.}}:
\batitle{Non-gaussian quantum state generation by multi-photon subtraction at the telecommunication wavelength}.
\bjtitle{Optics Express}
\bvolume{31}(\bissue{8}),
\bfpage{12865}
(\byear{2023})
\doiurl{10.1364/oe.486270}
\end{barticle}
\endbibitem

\bibitem[\protect\citeauthoryear{Yukawa et~al.}{2013}]{Yukawa2013}
\begin{barticle}
\bauthor{\bsnm{Yukawa}, \binits{M.}},
\bauthor{\bsnm{Miyata}, \binits{K.}},
\bauthor{\bsnm{Mizuta}, \binits{T.}},
\bauthor{\bsnm{Yonezawa}, \binits{H.}},
\bauthor{\bsnm{Marek}, \binits{P.}},
\bauthor{\bsnm{Filip}, \binits{R.}},
\bauthor{\bsnm{Furusawa}, \binits{A.}}:
\batitle{Generating superposition of up-to three photons for continuous variable quantum information processing}.
\bjtitle{Optics Express}
\bvolume{21}(\bissue{5}),
\bfpage{5529}--\blpage{35}
(\byear{2013})
\doiurl{10.1364/OE.21.005529}
\end{barticle}
\endbibitem

\bibitem[\protect\citeauthoryear{Mari and Eisert}{2012}]{Mari2012}
\begin{barticle}
\bauthor{\bsnm{Mari}, \binits{A.}},
\bauthor{\bsnm{Eisert}, \binits{J.}}:
\batitle{Positive wigner functions render classical simulation of quantum computation efficient}.
\bjtitle{Physical Review Letters}
\bvolume{109}(\bissue{23}),
\bfpage{230503}
(\byear{2012})
\doiurl{10.1103/PhysRevLett.109.230503}
\end{barticle}
\endbibitem

\bibitem[\protect\citeauthoryear{Kashiwazaki et~al.}{2023}]{Kashiwazaki2023}
\begin{botherref}
\oauthor{\bsnm{Kashiwazaki}, \binits{T.}},
\oauthor{\bsnm{Yamashima}, \binits{T.}},
\oauthor{\bsnm{Enbutsu}, \binits{K.}},
\oauthor{\bsnm{Kazama}, \binits{T.}},
\oauthor{\bsnm{Inoue}, \binits{A.}},
\oauthor{\bsnm{Fukui}, \binits{K.}},
\oauthor{\bsnm{Endo}, \binits{M.}},
\oauthor{\bsnm{Umeki}, \binits{T.}},
\oauthor{\bsnm{Furusawa}, \binits{A.}}:
Over-8-{dB} squeezed light generation by a broadband waveguide optical parametric amplifier toward fault-tolerant ultra-fast quantum computers.
Applied Physics Letters
\textbf{122}(23)
(2023)
\doiurl{10.1063/5.0144385}
\end{botherref}
\endbibitem

\bibitem[\protect\citeauthoryear{Hattori et~al.}{2022}]{Hattori2022}
\begin{barticle}
\bauthor{\bsnm{Hattori}, \binits{K.}},
\bauthor{\bsnm{Konno}, \binits{T.}},
\bauthor{\bsnm{Miura}, \binits{Y.}},
\bauthor{\bsnm{Takasu}, \binits{S.}},
\bauthor{\bsnm{Fukuda}, \binits{D.}}:
\batitle{An optical transition-edge sensor with high energy resolution}.
\bjtitle{Superconductor Science and Technology}
\bvolume{35}(\bissue{9}),
\bfpage{095002}
(\byear{2022})
\doiurl{10.1088/1361-6668/ac7e7b}
\end{barticle}
\endbibitem

\bibitem[\protect\citeauthoryear{Schrödinger}{1935}]{Schrodinger1935}
\begin{barticle}
\bauthor{\bsnm{Schrödinger}, \binits{E.}}:
\batitle{Die gegenwärtige situation in der quantenmechanik}.
\bjtitle{Die Naturwissenschaften}
\bvolume{23}(\bissue{50}),
\bfpage{844}--\blpage{849}
(\byear{1935})
\doiurl{10.1007/bf01491987}
\end{barticle}
\endbibitem

\bibitem[\protect\citeauthoryear{Endo et~al.}{2025}]{supplement}
\begin{botherref}
\oauthor{\bsnm{Endo}, \binits{M.}},
\oauthor{\bsnm{Nomura}, \binits{T.}},
\oauthor{\bsnm{Sonoyama}, \binits{T.}},
\oauthor{\bsnm{Takahashi}, \binits{K.}},
\oauthor{\bsnm{Takasu}, \binits{S.}},
\oauthor{\bsnm{Fukuda}, \binits{D.}},
\oauthor{\bsnm{Kashiwazaki}, \binits{T.}},
\oauthor{\bsnm{Inoue}, \binits{A.}},
\oauthor{\bsnm{Umeki}, \binits{T.}},
\oauthor{\bsnm{Nehra}, \binits{R.}},
\oauthor{\bsnm{Marek}, \binits{P.}},
\oauthor{\bsnm{Filip}, \binits{R.}},
\oauthor{\bsnm{Takase}, \binits{K.}},
\oauthor{\bsnm{Asavanant}, \binits{W.}},
\oauthor{\bsnm{Furusawa}, \binits{A.}}:
Supplementary Information
(2025)
\end{botherref}
\endbibitem

\bibitem[\protect\citeauthoryear{Lvovsky and Raymer}{2009}]{Lvovsky2009}
\begin{barticle}
\bauthor{\bsnm{Lvovsky}, \binits{A.I.}},
\bauthor{\bsnm{Raymer}, \binits{M.G.}}:
\batitle{Continuous-variable optical quantum-state tomography}.
\bjtitle{Reviews of Modern Physics}
\bvolume{81}(\bissue{1}),
\bfpage{299}--\blpage{332}
(\byear{2009})
\doiurl{10.1103/RevModPhys.81.299}
\end{barticle}
\endbibitem

\bibitem[\protect\citeauthoryear{Fedotova et~al.}{2023}]{Fedotova2023}
\begin{botherref}
\oauthor{\bsnm{Fedotova}, \binits{E.}},
\oauthor{\bsnm{Kuznetsov}, \binits{N.}},
\oauthor{\bsnm{Tiunov}, \binits{E.}},
\oauthor{\bsnm{Ulanov}, \binits{A.E.}},
\oauthor{\bsnm{Lvovsky}, \binits{A.I.}}:
Continuous-variable quantum tomography of high-amplitude states.
Physical Review A
\textbf{108}(4)
(2023)
\doiurl{10.1103/PhysRevA.108.042430}
\end{botherref}
\endbibitem

\bibitem[\protect\citeauthoryear{Killoran et~al.}{2019}]{Killoran2019}
\begin{botherref}
\oauthor{\bsnm{Killoran}, \binits{N.}},
\oauthor{\bsnm{Izaac}, \binits{J.}},
\oauthor{\bsnm{Quesada}, \binits{N.}},
\oauthor{\bsnm{Bergholm}, \binits{V.}},
\oauthor{\bsnm{Amy}, \binits{M.}},
\oauthor{\bsnm{Weedbrook}, \binits{C.}}:
Strawberry fields: A software platform for photonic quantum computing.
Quantum
\textbf{3}
(2019)
\doiurl{10.22331/q-2019-03-11-129}
\end{botherref}
\endbibitem

\bibitem[\protect\citeauthoryear{Bromley et~al.}{2020}]{Bromley2020}
\begin{botherref}
\oauthor{\bsnm{Bromley}, \binits{T.R.}},
\oauthor{\bsnm{Arrazola}, \binits{J.M.}},
\oauthor{\bsnm{Jahangiri}, \binits{S.}},
\oauthor{\bsnm{Izaac}, \binits{J.}},
\oauthor{\bsnm{Quesada}, \binits{N.}},
\oauthor{\bsnm{Gran}, \binits{A.D.}},
\oauthor{\bsnm{Schuld}, \binits{M.}},
\oauthor{\bsnm{Swinarton}, \binits{J.}},
\oauthor{\bsnm{Zabaneh}, \binits{Z.}},
\oauthor{\bsnm{Killoran}, \binits{N.}}:
Applications of near-term photonic quantum computers: software and algorithms.
Quantum Science and Technology
\textbf{5}(3)
(2020)
\doiurl{10.1088/2058-9565/ab8504}
\end{botherref}
\endbibitem

\bibitem[\protect\citeauthoryear{Parigi et~al.}{2007}]{Parigi2007}
\begin{barticle}
\bauthor{\bsnm{Parigi}, \binits{V.}},
\bauthor{\bsnm{Zavatta}, \binits{A.}},
\bauthor{\bsnm{Kim}, \binits{M.}},
\bauthor{\bsnm{Bellini}, \binits{M.}}:
\batitle{Probing quantum commutation rules by addition and subtraction of single photons to/from a light field}.
\bjtitle{Science}
\bvolume{317}(\bissue{5846}),
\bfpage{1890}--\blpage{3}
(\byear{2007})
\doiurl{10.1126/science.1146204}
\end{barticle}
\endbibitem

\bibitem[\protect\citeauthoryear{Chen et~al.}{2024}]{Chen2024}
\begin{botherref}
\oauthor{\bsnm{Chen}, \binits{Y.-R.}},
\oauthor{\bsnm{Hsieh}, \binits{H.-Y.}},
\oauthor{\bsnm{Ning}, \binits{J.}},
\oauthor{\bsnm{Wu}, \binits{H.-C.}},
\oauthor{\bsnm{Chen}, \binits{H.L.}},
\oauthor{\bsnm{Shi}, \binits{Z.-H.}},
\oauthor{\bsnm{Yang}, \binits{P.}},
\oauthor{\bsnm{Steuernagel}, \binits{O.}},
\oauthor{\bsnm{Wu}, \binits{C.-M.}},
\oauthor{\bsnm{Lee}, \binits{R.-K.}}:
Generation of heralded optical cat states by photon addition.
Physical Review A
\textbf{110}(2)
(2024)
\doiurl{10.1103/PhysRevA.110.023703}
\end{botherref}
\endbibitem

\bibitem[\protect\citeauthoryear{Fadrný et~al.}{2024}]{Fadrny2024}
\begin{botherref}
\oauthor{\bsnm{Fadrný}, \binits{J.}},
\oauthor{\bsnm{Neset}, \binits{M.}},
\oauthor{\bsnm{Bielak}, \binits{M.}},
\oauthor{\bsnm{Ježek}, \binits{M.}},
\oauthor{\bsnm{Bílek}, \binits{J.}},
\oauthor{\bsnm{Fiurášek}, \binits{J.}}:
Experimental preparation of multiphoton-added coherent states of light.
npj Quantum Information
\textbf{10}(1)
(2024)
\doiurl{10.1038/s41534-024-00885-y}
\end{botherref}
\endbibitem

\bibitem[\protect\citeauthoryear{Su et~al.}{2019}]{Su2019}
\begin{barticle}
\bauthor{\bsnm{Su}, \binits{D.}},
\bauthor{\bsnm{Myers}, \binits{C.R.}},
\bauthor{\bsnm{Sabapathy}, \binits{K.K.}}:
\batitle{Conversion of gaussian states to non-gaussian states using photon-number-resolving detectors}.
\bjtitle{Physical Review A}
\bvolume{100}(\bissue{5}),
\bfpage{052301}
(\byear{2019})
\doiurl{10.1103/PhysRevA.100.052301}
\end{barticle}
\endbibitem

\bibitem[\protect\citeauthoryear{Umeki et~al.}{2010}]{Umeki2010}
\begin{barticle}
\bauthor{\bsnm{Umeki}, \binits{T.}},
\bauthor{\bsnm{Tadanaga}, \binits{O.}},
\bauthor{\bsnm{Asobe}, \binits{M.}}:
\batitle{Highly efficient wavelength converter using direct-bonded ppznln ridge waveguide}.
\bjtitle{IEEE Journal of Quantum Electronics}
\bvolume{46}(\bissue{8}),
\bfpage{1206}--\blpage{1213}
(\byear{2010})
\doiurl{10.1109/jqe.2010.2045475}
\end{barticle}
\endbibitem

\bibitem[\protect\citeauthoryear{Kashiwazaki et~al.}{2021}]{Kashiwazaki2021}
\begin{barticle}
\bauthor{\bsnm{Kashiwazaki}, \binits{T.}},
\bauthor{\bsnm{Yamashima}, \binits{T.}},
\bauthor{\bsnm{Takanashi}, \binits{N.}},
\bauthor{\bsnm{Inoue}, \binits{A.}},
\bauthor{\bsnm{Umeki}, \binits{T.}},
\bauthor{\bsnm{Furusawa}, \binits{A.}}:
\batitle{Fabrication of low-loss quasi-single-mode ppln waveguide and its application to a modularized broadband high-level squeezer}.
\bjtitle{Applied Physics Letters}
\bvolume{119}(\bissue{25}),
\bfpage{251104}
(\byear{2021})
\doiurl{10.1063/5.0063118}
\end{barticle}
\endbibitem

\end{thebibliography}

\end{document}



\title{Supplementary Information for\\ ``High-Rate Four Photon Subtraction from Squeezed Vacuum: Preparing Cat State for Optical Quantum Computation''}

\date{\today}

\maketitle

\setcounter{equation}{0}
\renewcommand{\theequation}{S\arabic{equation}}

\setcounter{figure}{0}
\renewcommand{\thefigure}{S\arabic{figure}}

\setcounter{table}{0}
\renewcommand{\thetable}{S\arabic{table}}

\section{Details of experimental apparatus}\label{sec:apparatus}

\subsection{Pump, probe, and LO beam preparation}
\begin{figure*}
\centering
\includegraphics[width=\textwidth]{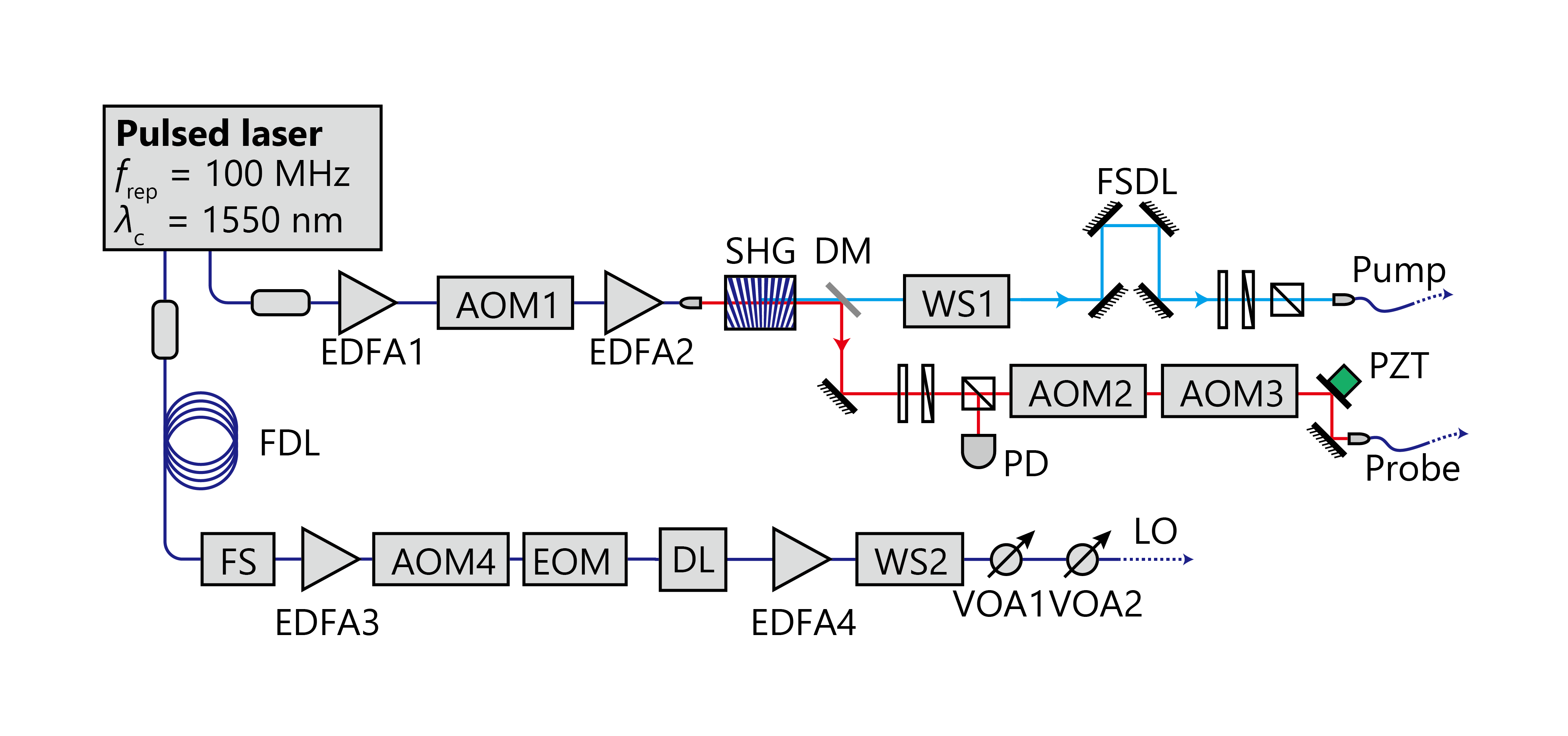}
\caption{Setup for the pump, probe, and LO light preparation.}\label{lights}
\end{figure*}

As shown in Fig.~\ref{lights}, a femtosecond mode-locked fiber laser (M-Comb-ULN, Menlo Systems) with a center wavelength of \SI{1550}{nm}, a spectral bandwidth of over \SI{40}{nm}, and a repetition rate of \SI{100}{MHz} was employed as the light source in this experiment. Since the timing resolution of the TES used in our experiment was limited to approximately \SI{50}{ns}, we reduced the pulse repetition rate to \SI{5}{MHz} via pulse picking with AOMs, and then applied bandpass filters to restrict the optical spectrum, thereby preparing a non-Gaussian state with a temporal width of about \SI{10}{ps}. Likewise, the probe light for monitoring the squeezed light phase during the parametric process and the LO light for homodyne detection were also pulse-picked and spectrally filtered. Details are provided below.

\subsubsection{Pump and probe lights}
One of the multiple output ports of the mode-locked laser was used to generate both the pump and probe lights. All optical fibers used in this experiment were polarization-maintaining (PM) fibers. First, the spectrum of the output light was limited by an optical bandpass filter to a center wavelength of \SI{1545.32}{nm} and a spectral width of \SI{1.0}{nm}, and it was then amplified by a homemade erbium-doped fiber amplifier (EDFA1) using an Er80-8/125-PM fiber (nLight) as the gain medium. Next, a fiber-coupled AOM (AOM1; SGTF400-1550-1P(H), Chongqing Smart Science \& Technology Development) driven at \SI{400}{MHz} was used for pulse picking, thus lowering the repetition rate to \SI{5}{MHz}. The driving and gating signals for this AOM were obtained by frequency dividing the repetition rate of the mode-locked laser \cite{de-Vries2015}. The pulses were subsequently amplified again by EDFA2 (HPP-PMFA-22, PriTel) and then emitted into free space. The beam was then focused into a fan-out periodically-poled MgO-doped near-stoichiometric \(\text{LiTaO}_3\) crystal (fan-out PPMgSLT, OXIDE), generating second-harmonic light at a center wavelength of \SI{772.66}{nm} and a spectral width of \SI{0.5}{nm}. The second-harmonic and fundamental lights were separated by a dichroic mirror (DM) and used as the pump and probe beams, respectively. To suppress unwanted spectral components, we applied a homemade wave shaper (WS1). To match their timing, a free-space delay line (FSDL) was used for temporal adjustment, followed by a half-wave plate (HWP) and a polarization beam splitter (PBS) to adjust the pump power. Subsequently, the pump light was coupled into a PM single-mode fiber. Part of the fundamental light was split off by a HWP and a PBS and directed to a photodiode (PD) for trigger signals, while the remaining fundamental light was used as the probe beam. The probe beam was modulated by two free-space AOMs (AOM2, AOM3; AOM3080-1912, Gooch \& Housego) at \SI{500}{Hz} for the sample-and-hold phase locking method. In addition, for heterodyne beat locking, AOM2 and AOM3 were driven at \SI{-80.0}{MHz} and \SI{80.1}{MHz}, respectively, creating a difference frequency of \SI{100}{kHz}. A mirror with a piezoelectric transducer (PZT) was placed in the beam path for phase locking, and the beam was then coupled into a PM fiber.

\subsubsection{LO light}
The LO beam was obtained from a different output port of the mode-locked laser. In most pulsed-light experiments, the LO and pump beams are generated from the same port \cite{Ourjoumtsev2007,Gerrits2010,Namekata2010,Bouillard2019}; however, in that case the squeezed pulses after the OPA typically exhibit a shorter duration and broader bandwidth than the LO \cite{Gerrits2010,Eto2011} and experience dispersive effects in the SHG and OPA processes \cite{Taguchi2020}, making it difficult to realize high-level squeezing and large Wigner negativity without pulse shaping \cite{Gerrits2010,Eto2011}. Here, we used light taken from a separate output port and employed a commercial waveform shaper (WS2; WaveShaper 1000A, II-VI) to achieve more flexible waveform control of the LO beam. Specifically, the spectrum of the output beam was restricted by an optical bandpass filter (center wavelength: \SI{1545.32}{nm}, bandwidth: \SI{12}{nm}), and then the pulse was pre-chirped (stretched) using a \SI{32}{m} fiber delay line (FDL) to mitigate distortion caused by subsequent fiber amplification. A fiber stretcher (FS) using a cylindrical piezoelectric transducer (PT140.70, PI) was also inserted for long-term phase stabilization. The beam was then amplified by EDFA3 (EDFA100P, Thorlabs), after which it was pulse-picked by a fiber-coupled AOM (AOM4; SGTF400-1550-1P(H), Chongqing Smart Science \& Technology Development) to reduce the repetition rate to \SI{5}{MHz}. Subsequently, it was sent through a waveguide phase modulator (EOM; MPX-LN-0.1-00-P-P-FA-FA, iXblue) for phase stabilization in the feedback control loop. An optical delay line (DL; ODL-600-11-1550-8/125-P-60-3A3A-1-1-600, OZ optics) was used to fine-tune the optical path length. The beam was amplified again by EDFA4 (EDFA100P, Thorlabs), and then the spectral amplitude and phase were controlled by the WS2. Finally, the beam passed through variable optical attenuators (VOA1, MMVOA-1-1550-P-8/125-3A3A-1-0.5, OZ optics; VOA2, NVOA-325125333-D0601, Agiltron) to stabilize and modulate the LO power, and was emitted as free-space light through a collimator for homodyne detection. In this experiment, the WS2 was programmed to produce a Gaussian waveform at a center wavelength of \SI{1545.32}{nm} with a spectral width of \SI{0.8}{nm}, to which a second-order dispersion of about \SI{-0.7}{ps/nm} was added to compensate for dispersion introduced by the optical elements. Autocorrelation measurements revealed an \SI{8}{ps} pulse duration, corresponding to the transform-limited pulse width.

\section{Measured quadrature data}
\begin{figure*}
\centering
\includegraphics[width=\textwidth]{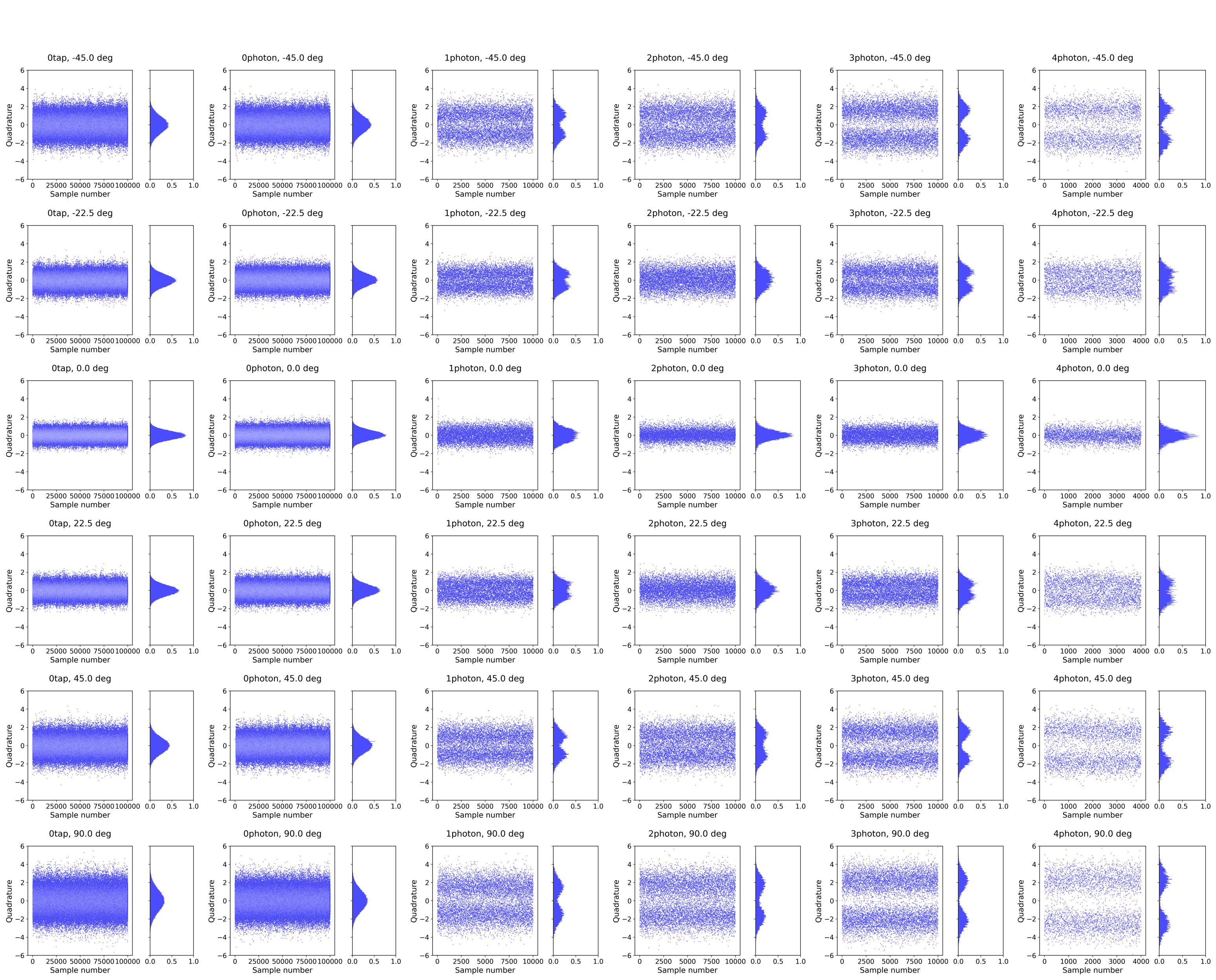}
\caption{Scatter plots and histograms of the measured quadrature data.}\label{quadratures}
\end{figure*}

We summarized the measured quadrature-phase amplitudes obtained by homodyne detection into scatter plots and histograms, as shown in Fig.~\ref{quadratures}. Based on these results, we estimated the density matrix of the quantum state via a maximum-likelihood method without applying any loss correction. The error bars were calculated using 1000 iterations of the bootstrap method.

\section{Density matrices}
\begin{figure*}
\centering
\includegraphics[width=\textwidth]{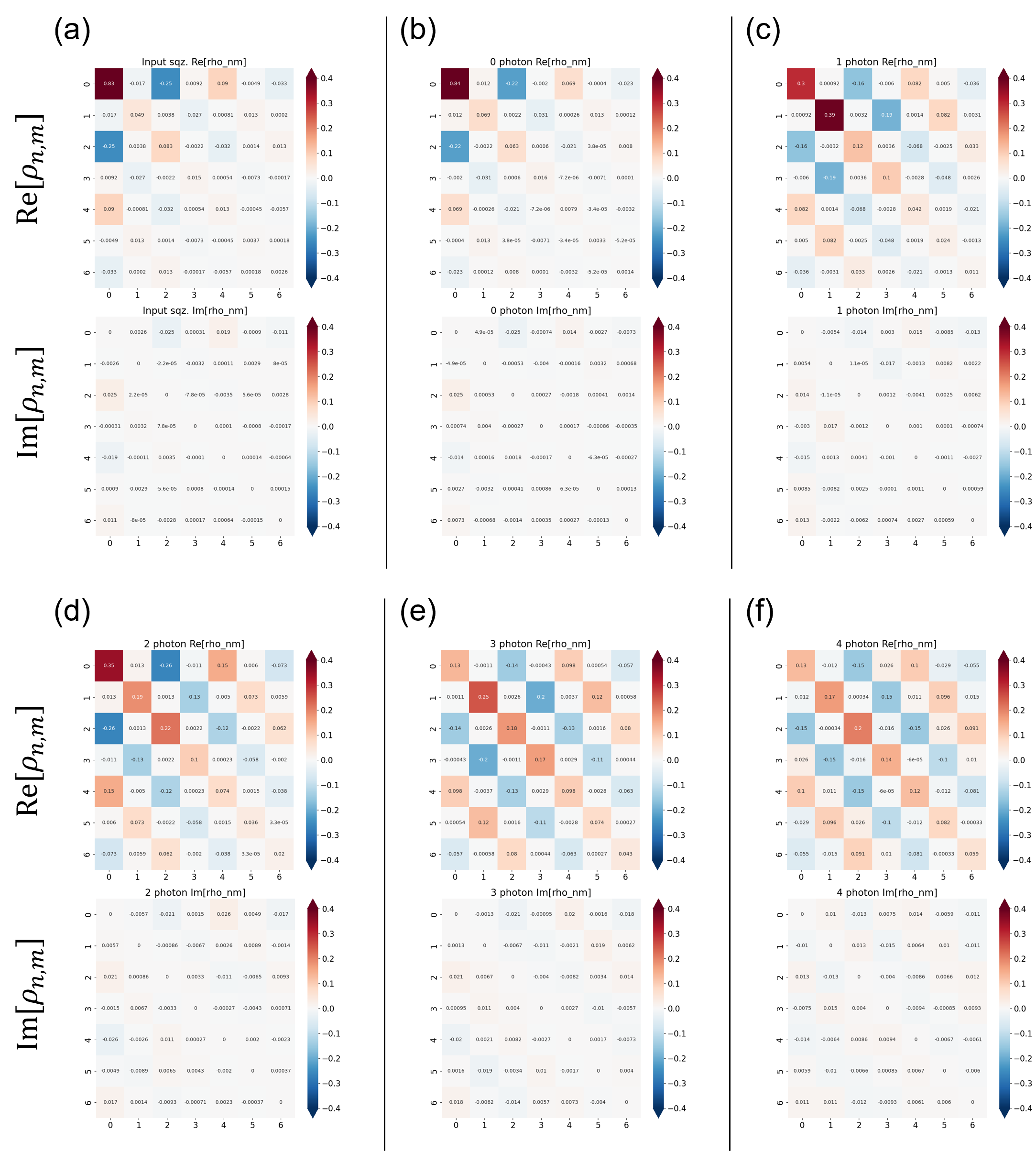}
\caption{Experimentally obtained density matrices in the photon number basis (\(\mathrm{Re}[\rho_{n,m}], \mathrm{Im}[\rho_{n,m}]\)). (a) Input squeezed vacuum and (b-f) zero- to four-photon subtraction, respectively.}\label{rhos_nm}
\end{figure*}

\begin{figure*}
\centering
\includegraphics[width=\textwidth]{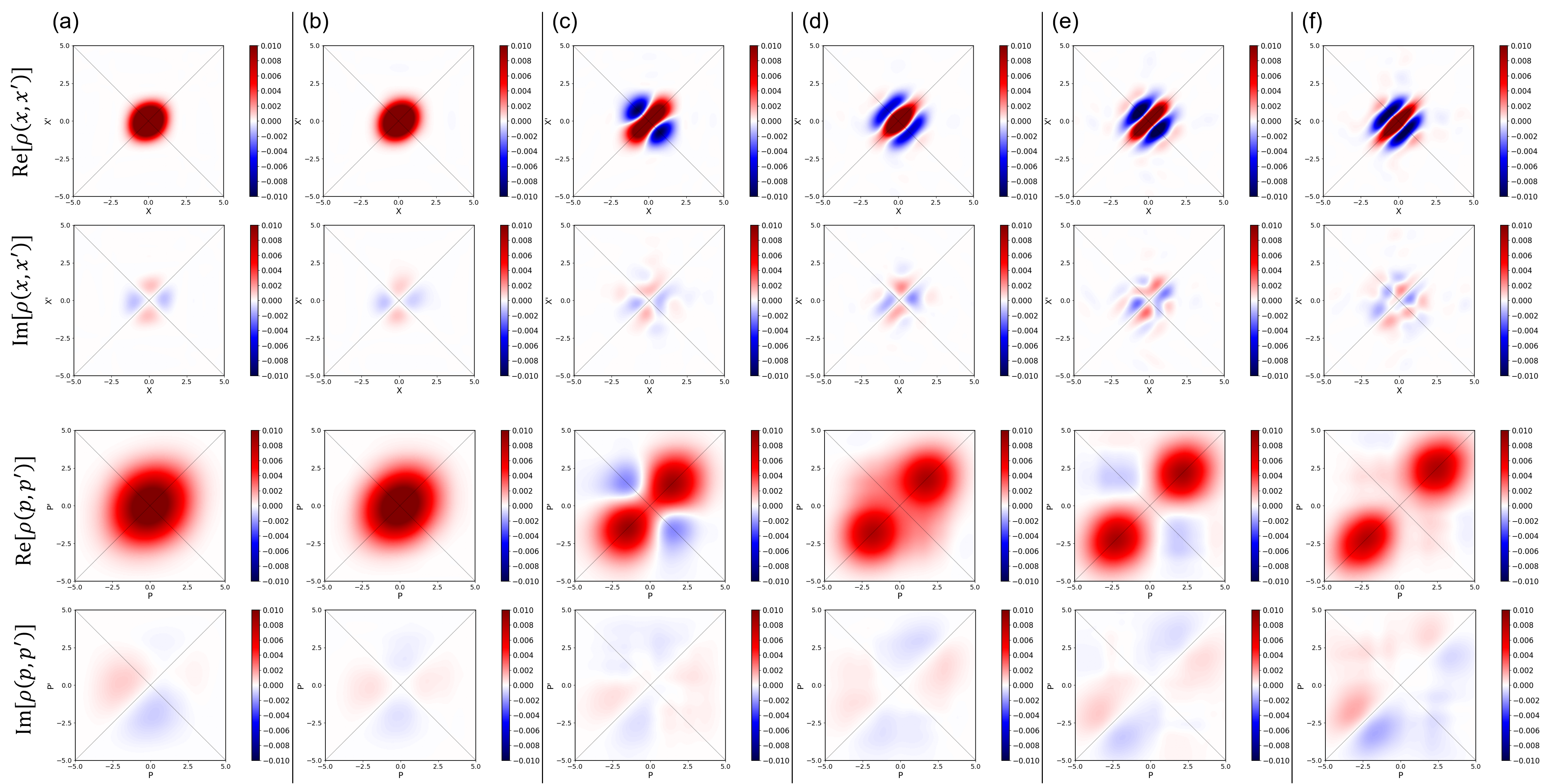}
\caption{Experimentally obtained density matrices in the position and momentum bases (\(\mathrm{Re}[\rho(x,x')],\mathrm{Im}[\rho(x,x')],\mathrm{Re}[\rho(p,p')],\mathrm{Im}[\rho(p,p')]\)). (a) Input squeezed vacuum and (b-f) zero- to four-photon subtraction, respectively.}\label{rhos_xx_pp}
\end{figure*}

Experimentally obtained density matrices are shown in Fig.~\ref{rhos_nm} and Fig.~\ref{rhos_xx_pp}. Note that the real parts of the density matrices in the momentum basis (\(\mathrm{Re}[\rho(p,p')]\)) are identical to those presented in the main text.

\section{Marginal distributions}
\begin{figure*}
\centering
\includegraphics[width=\textwidth]{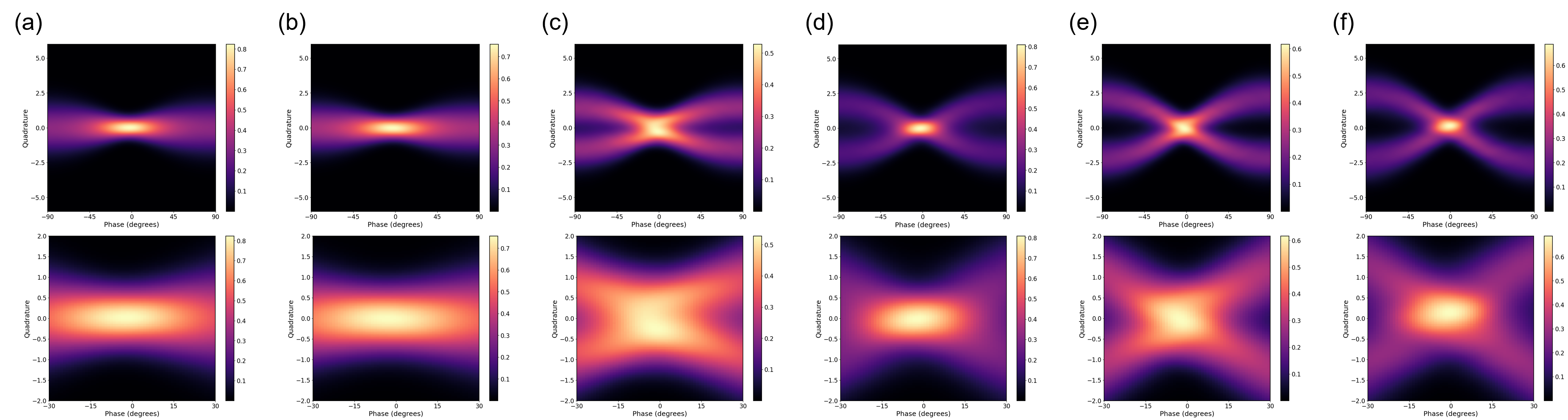}
\caption{Calculated marginal distributions from the experimental data. (a) Input squeezed vacuum and (b-f) zero- to four-photon subtraction, respectively. The upper plot displays data ranging from -90° to 90°, while the lower plot presents a magnified view of the range from -30° to 30°.}\label{marginals}
\end{figure*}

Figure~\ref{marginals} shows the marginal distributions calculated from the experimental data.

%